%% file: main.tex
\documentclass[sigconf, nonacm]{acmart}
\usepackage{amsmath,amsfonts}
\usepackage{algorithmic}
\usepackage{graphicx}
\usepackage{textcomp}
\usepackage{xcolor}
\usepackage{booktabs}
\usepackage{xcolor}
\usepackage{pgfplots}
\usepackage{subcaption}
\usepackage{colortbl}
\usepackage{braket}
\usepackage{epsfig,latexsym,graphics}
\usepackage[font={small}]{caption}
\usepackage{setspace}
\usepackage{multirow}
\usepackage{enumitem}
\pgfplotsset{compat=1.16}
\usepackage{cmd}
\usepackage[spacesep,definevectors]{easyvector} 
\usepackage[a-2b]{pdfx}
\newcommand\vldbdoi{XX.XX/XXX.XX}
\newcommand\vldbpages{XXX-XXX}
\newcommand\vldbvolume{19}
\newcommand\vldbissue{6}
\newcommand\vldbyear{2026}
\newcommand\vldbauthors{\authors}
\newcommand\vldbtitle{\shorttitle} 
\newcommand\vldbavailabilityurl{https://github.com/mingyu-hkustgz/RESQ}
\newcommand\vldbpagestyle{plain} 

\begin{document}
\title{Quantization Meets Projection: A Happy Marriage for Approximate k-Nearest Neighbor Search}

\author{Mingyu Yang}
\affiliation{%
  \institution{HKUST (GZ) \& HKUST}
}
\email{myang250@connect.hkust-gz.edu.cn}

\author{Liuchang Jing}
\affiliation{%
  \institution{HKUST (GZ)}
}
\email{ljing248@connect.hkust-gz.edu.cn}

\author{Wentao Li}
\affiliation{%
  \institution{University of Leicester}
}
\email{wl226@leicester.ac.uk}

\author{Wei Wang}
\affiliation{%
  \institution{HKUST (GZ) \& HKUST}
}
\email{weiwcs@ust.hk}

\begin{abstract}
Approximate $k$-nearest neighbor (AKNN) search is a fundamental problem with wide applications.
To reduce memory and accelerate search, vector quantization is widely adopted.
However, existing quantization methods either rely on codebooks---whose query speed is limited by costly table lookups---or adopt dimension-wise quantization, which maps each vector dimension to a small quantized code for fast search. 
The latter, however, suffers from a fixed compression ratio because the quantized code length is inherently tied to the original dimensionality.
To overcome these limitations, we propose $\MRQ$, a new approach that integrates projection with quantization.
The key insight is that, after projection, high-dimensional vectors tend to concentrate most of their information in the leading dimensions.
$\MRQ$ exploits this property by quantizing only the information-dense projected subspace---whose size is fully user-tunable---thereby decoupling the quantized code length from the original dimensionality.
The remaining tail dimensions are captured using lightweight statistical summaries.
By doing so, $\MRQ$ boosts the query efficiency of existing quantization methods while achieving arbitrary compression ratios enabled by the projection step.
Extensive experiments show that $\MRQ$ substantially outperforms the state-of-the-art method, achieving up to 3× faster search with only one-third the quantization bits for comparable accuracy.
\end{abstract}

\maketitle

\pagestyle{\vldbpagestyle}
\begingroup\small\noindent\raggedright\textbf{PVLDB Reference Format:}\\
\vldbauthors. \vldbtitle. PVLDB, \vldbvolume(\vldbissue): \vldbpages, \vldbyear.\\
\href{https://doi.org/\vldbdoi}{doi:\vldbdoi}
\endgroup
\begingroup
\renewcommand\thefootnote{}\footnote{\noindent
This work is licensed under the Creative Commons BY-NC-ND 4.0 International License. Visit \url{https://creativecommons.org/licenses/by-nc-nd/4.0/} to view a copy of this license. For any use beyond those covered by this license, obtain permission by emailing \href{mailto:info@vldb.org}{info@vldb.org}. Copyright is held by the owner/author(s). Publication rights licensed to the VLDB Endowment. \\
\raggedright Proceedings of the VLDB Endowment, Vol. \vldbvolume, No. \vldbissue\ %
ISSN 2150-8097. \\
\href{https://doi.org/\vldbdoi}{doi:\vldbdoi} \\
}\addtocounter{footnote}{-1}\endgroup

\ifdefempty{\vldbavailabilityurl}{}{
\vspace{.3cm}
\begingroup\small\noindent\raggedright\textbf{PVLDB Artifact Availability:}\\
The source code, data, and/or other artifacts have been made available at \url{\vldbavailabilityurl}.
\endgroup
}

\section{Introduction}
The $k$-Nearest Neighbor (KNN) search for vectors is critical to many applications, including recommendation systems~\cite{CF-2007-recommender-sys}, information retrieval~\cite{Image-Search-2007-PR}, and Retrieval-Augmented Generation for Large Language Models~\cite{LLM-RAG-NIPS-2020}.
However, the curse of dimensionality~\cite{Curse-of-dim-1998} renders exact KNN search computationally expensive in high-dimensional space.
This has led to growing interest in approximate KNN (AKNN) search~\cite{DBLP:journals/pvldb/NSGFuXWC19,DBLP:journals/pami/SSGFuWC22,SRS-yifang-2014,PMLSH-bolong-2020,ADSampling:journals/sigmod/GaoL23,Graph-ANNS-Survey-VLDB-2021-mengzhao,C2LSH-2012-SIGMOD,LSH-Pstable-2004,LSH-1999,QALSH-2015-VLDB,FANNG:harwood2016fanng,Finger-WWW-2023,VSAG,DEG-SIGMOD-2025,SymphonyQG-SIGMOD-2025,SeRF-SIGMOD-2024,DIGRA-SIGMOD-2025,RangePQ-SIGMOD-2025,UNG-SIGMOD-2024,ACORN-SIGMOD-2023}, which trades off a small amount of accuracy for substantially improved efficiency in large-scale settings.

\stitle{Quantization.}
The vectors used in AKNN search are typically stored in floating-point format, making large-scale in-memory storage expensive. To mitigate this, \textbf{quantization} has become a widely adopted technique for reducing storage overhead~\cite{VQ-IEEE=1984,VQ-image-1998-TOC}. Prominent quantization methods include Binary Quantization (BQ)~\cite{ITQ:/pami/GongLGP13,Rabitq-SIGMOD-2024,IsoHash-NIPS-2012,Binary-LSH-NIPS-2009,GQR-learn-to-hash-SIGMOD-2018}, Scalar Quantization (SQ)~\cite{LVQ-2024-VLDB,SQ-IR-MM-2012,ExRaBitQ-arxiv-2024}, Product Quantization (PQ)~\cite{PQ-fast-scan-VLDB-2015,LOPQ-2014-CVPR,DBLP:journals/pami/OPQGeHK014,AQ-2014-cvpr,Quicker-adc-PAMI-2019,DeltaPQ-VLDB-2020,LVQ-2024-VLDB,Route-Guide-PQ-ICDE-2024,Norm-Exp-Q-AAAI-2020,LSQ++-2018-eccv}, and Additive Quantization (AQ)~\cite{AQ-2014-cvpr,CompQ-2014-ICML,TreeQ-CVPR-2015,RQ-TIP-1996}. 
These methods can be broadly grouped into two families based on their encoding strategies.
\noindent\textbf{(1) Codebook-Based Quantization.} Methods such as PQ, AQ, and their variants~\cite{CompQ-2014-ICML,TreeQ-CVPR-2015,DeltaPQ-VLDB-2020} learn a set of representative vectors (codewords), collectively referred to as a \emph{codebook}. Each original vector is then approximately represented as a composition of codewords. These methods offer flexible compression ratios through configurable codebook sizes. 
Yet, they rely on lookup tables and codeword aggregation for distance computation during AKNN search, which can become a computational bottleneck without specialized optimizations~\cite{PQ-fast-scan-VLDB-2015}.
\noindent\textbf{(2) Dimension-Wise Quantization.} Methods such as BQ and SQ quantize each vector dimension independently. For example, BQ converts each dimension to a single bit, while SQ transforms the numeric format of each dimension (e.g., from a 32-bit float to an 8-bit or 4-bit integer). 
These methods achieve very high query efficiency as they avoid codeword lookups required by codebook-based approaches.

\stitle{Motivations.}
As confirmed by recent studies, dimension-wise quantization methods significantly outperform codebook-based approaches in both efficiency and accuracy~\cite{Rabitq-SIGMOD-2024,ExRaBitQ-arxiv-2024,LVQ-2024-VLDB,LVQ-VLDB-2023}.
Owing to their strong empirical performance, they have been widely deployed in large-scale data science applications and systems~\cite{Chameleon-VLDB-2025-SDS,VSAG,faiss:johnson2019billion,ES-2024-System}.
Their speed advantage primarily arises from their Single Instruction, Multiple Data (SIMD)–friendly design---leveraging low-level instructions such as \texttt{POPCNT} and \texttt{AVX-int8}---as well as beneficial geometric properties like unbiased inner-product estimation.

However, a key limitation of dimension-wise quantization is its restricted compression ratio.
Because these methods quantize each original dimension independently, the length of the quantized code remains tied to the original vector dimensionality.
For example, the state-of-the-art dimension-wise method RabitQ~\cite{Rabitq-SIGMOD-2024} requires the length of the quantized code to exactly match the dimensionality of the original vector.
This inflexibility introduces two major issues: 
(1) users cannot achieve higher compression ratios, which is problematic for large-scale datasets that may still exceed main-memory capacity even after quantization; and 
(2) users cannot reduce the number of quantization bits to improve efficiency. 
In addition, a fixed code length may misalign with SIMD bit-widths, leading to suboptimal performance.

\stitle{Our Solution.}
To summarize, existing quantization methods either suffer from slow query performance or impose a fixed compression ratio as the code length is tied to the original dimensionality. 
This raises a natural question: \emph{Can we retain high-performance quantization while avoiding the constraints of a fixed compression ratio?} 
We argue that a principled integration of projection and quantization provides a compelling answer.
Our key insight is grounded in an empirical analysis of modern high-dimensional vector embeddings: after applying an information-aware projection such as PCA, the per-dimension variance exhibits a pronounced long-tailed distribution (see Fig.~\ref{fig:vec_var_3}). This reveals that most information is concentrated within a low-dimensional \textit{subspace}, motivating differentiated treatment of projected dimensions during quantization.

Building on this observation, we propose our method $\MRQ$ (Minimized Residual Quantization).
$\MRQ$ first projects the original vector into a predefined number of dimensions, forming a \textbf{projected vector}, while the remaining dimensions constitute the \textbf{residual vector}.
For the projected vector---which captures most of the variance---we apply RabitQ~\cite{Rabitq-SIGMOD-2024}-style quantization, but extended to support \textit{arbitrary} bit lengths.
For the residual vector, we introduce a key innovation: instead of storing it directly, we model its distance contribution using lightweight statistics (e.g., variance). 
This enables accurate distance estimation with bounded error, further prunes the residual dimension during distance computation.
By combining the quantized projected vector with the residual component, $\MRQ$ performs multi-stage distance computation to ensure efficient and accurate AKNN search.
The process starts with fast, lightweight estimation and progressively refines the result using more accurate, but costlier, computations.
Fig.~1 illustrates this process. 
Crucially, $\MRQ$ supports an arbitrary compression ratio for quantization---unlike prior work---while still ensuring accuracy through the use of error bounds for re-ranking.

\stitle{Contributions.}
Our main contributions are summarized as follows:

\sstitle{A Flexible Quantization Mechanism.}
To address the limitation of the dimension-wise quantization methods (especially the state-of-the-art method RabitQ~\cite{Rabitq-SIGMOD-2024}), which enforces a fixed compression ratio, we are the first to systematically analyze and exploit the long-tailed variance distribution of high-dimensional data after projection rotation. This insight underpins a novel AKNN search method that seamlessly integrates projection and quantization.

\sstitle{A Novel Distance Computation Framework.}
We design a multi-stage distance computation framework that asymmetrically handles projected and residual components (see Fig.~1). It combines distance estimation and error bounds derived from both components for efficient and accurate AKNN search.

\sstitle{Efficient Implementation.}
We incorporate our multi-stage distance computation into a highly optimized IVF-based index. This design leverages custom memory layouts and approximation strategies to improve cache locality and dramatically reduce index construction time, ensuring scalability to large-scale datasets.

\sstitle{Extensive Experimental Evaluation.}
We conduct comprehensive evaluations on real-world datasets, benchmarking against strong baselines including the graph-based method HNSW and the quantization-based method RabitQ. 
Our method consistently outperforms existing approaches, achieving up to a $3\times$ improvement in search efficiency while preserving comparable accuracy.

Due to space limitations, some discussions and experiments are omitted here and can be found in the appendix.

\begin{figure}[!t]
    \centering
    \includegraphics[width=0.95\linewidth]{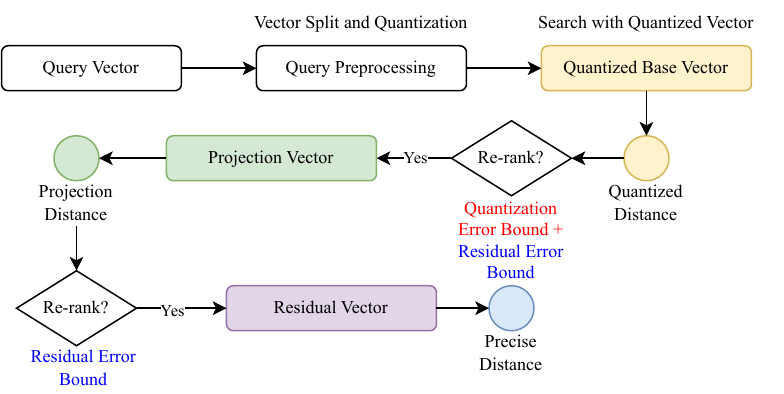}
    \caption{Workflow of $\MRQ$. After projection, both base and query vectors are split into projected and residual components.
$\MRQ$ performs distance computation in three stages:
(1) The projected vectors are quantized with arbitrary bit lengths, enabling fast distance estimation with error bounds for re-ranking.
(2) For higher accuracy, the original projected vectors replace the quantized ones, yielding more precise distances while retaining the error bounds.
(3) If exact computation is needed, the full vector (projected + residual) is used.}
    \label{fig:Multi-stage}
\end{figure}

\section{Preliminary and Related Work}\label{sec:preliminary}
Section~\ref{sub:aknn} introduces the AKNN search problem and reviews existing AKNN methods.
Section~\ref{sub:quan} presents vector quantization techniques for reducing space overhead.
Section~\ref{sub:re-rank} describes re-ranking strategies used to improve search accuracy.

\subsection{AKNN Search and Methods}\label{sub:aknn}
Given a database $S$ of $n$ vectors/points in $D$-dimensional Euclidean space $\mathbb{R}^D$ and a query point $q \in \mathbb{R}^D$, the problem of KNN search is to retrieve the top $k$ points in $S$ closest to $q$ under the Euclidean distance. 
Due to the prohibitive cost of exact KNN search in high-dimensional spaces~\cite{Curse-of-dim-1998}, recent efforts have focused on approximate KNN (AKNN) methods. 
A common metric for evaluating AKNN accuracy is Recall@K, defined as the proportion of true nearest neighbors successfully retrieved:
$\text{Recall}@k = \frac{|T \cap G|}{|G|}$
where $T$ is the set returned by the AKNN method, and $G$ is the set of ground-truth KNN for the query $q$.

The goal of AKNN methods is to achieve high recall with minimal computational cost.
In the past decades, a variety of AKNN methods have been proposed, which can be broadly categorized into four classes:
(1) hashing-based methods~\cite{learning-to-hash-survy-2017,QALSH-2015-VLDB,C2LSH-2012-SIGMOD,SRS-yifang-2014,PMLSH-bolong-2020,LSH-1999,LSH-Pstable-2004,RPLSH-1995,query-aware-LSH,NeuralLSH-2020-ICLR,ADSampling:journals/sigmod/GaoL23},
(2) inverted-file (IVF)-based methods~\cite{DeltaPQ-VLDB-2020,PQ-fast-scan-VLDB-2015,LOPQ-2014-CVPR,DBLP:journals/pami/OPQGeHK014,CompQ-2014-ICML,AQ-2014-cvpr,Rabitq-SIGMOD-2024,ITQ:/pami/GongLGP13,IP-PQ-2016-guoruiqi,IMI:DBLP:journals/pami/BabenkoL15},
(3) tree-based methods~\cite{Cover-Tree-2006-ICML,Random-Project-Tree-2008,Herclude-VLDB-2022,ELPIS-VLDB-2023}, and
(4) graph-based methods~\cite{FANNG:harwood2016fanng,Finger-WWW-2023,NSWDBLP:journals/is/MalkovPLK14,DBLP:journals/pami/HNSWMalkovY20,DBLP:journals/pvldb/NSGFuXWC19,DBLP:journals/pami/SSGFuWC22,tMRNG:journals/pacmmod/PengCCYX23,Diskann-NIPS-2019,HVS:DBLP:journals/pvldb/LuKXI21,ELPIS-VLDB-2023,Accelerate-Graph-Index-Mengzhao-SIGMOD-2025,Graph-ANNS-Survey-VLDB-2021-mengzhao,Graph-Revisited-Jiadong-Xie-TODS2025,ELPIS-VLDB-2023}.
This paper focuses on IVF- and graph-based methods, which build specialized indexes to accelerate the search and achieve state-of-the-art performance.

\sstitle{IVF-based methods} first partition points in database $S$ using clustering algorithms such as $k$-means to obtain cluster centroids. 
Each data point is then assigned to its nearest centroid.
During query, we identify a small number of centroids closest to the query and search only the associated clusters to retrieve the top-$k$ results.

\sstitle{Graph-based methods} construct an index by modeling the high-dimensional data points as nodes in a graph, where edges connect each node to its nearby neighbors. This results in a navigable small-world graph structure.
At query time, the search starts from an entry node and iteratively traverses toward nodes that are progressively closer to the query.

To reduce space cost, these methods often adopt quantization techniques, as storing original float-type vectors is memory-intensive.
Also, re-ranking strategies are employed to improve search accuracy, since distance computations based on quantized vectors are inherently approximate.

\subsection{Quantization Techniques}\label{sub:quan}
Quantization compresses high-dimensional vectors into compact codes to reduce storage cost and improve search efficiency. 
Based on their encoding strategy, existing methods can be broadly categorized into \textit{codebook-based} and \textit{dimension-wise} approaches.

\stitle{Codebook-Based Quantization.} 
These methods approximate vectors using a finite set of representative centroids (i.e., codewords). 
For example, PQ~\cite{PQ-fast-scan-VLDB-2015,DeltaPQ-VLDB-2020,IP-PQ-2016-guoruiqi} decomposes a high-dimensional space into multiple low-dimensional subspaces and performs clustering within each subspace. 
Variants such as Optimized PQ (OPQ)~\cite{LOPQ-2014-CVPR,DBLP:journals/pami/OPQGeHK014} and Additive Quantization (AQ)~\cite{AQ-2014-cvpr,RQ-TIP-1996} further reduce quantization error through orthogonal rotations or additive code composition. 
These methods offer flexible compression ratios (controlled by the number of subspaces and codebook size). 
However, they suffer from a fundamental efficiency bottleneck: distance computation involves table lookups (random memory access) and floating-point additions, which are difficult to fully accelerate using SIMD instructions compared with bitwise operations. 
Although hardware-aware optimizations (e.g., 4-bit PQ with AVX-512~\cite{PQ-fast-scan-VLDB-2015}) improve performance, they typically impose strict constraints on codebook sizes or incur severe accuracy degradation~\cite{Rabitq-SIGMOD-2024}.


\stitle{Dimension-Wise Quantization.}
These methods quantize each dimension of a vector independently, mapping continuous values to discrete integers or bits. 
For example, Scalar Quantization (SQ)~\cite{LVQ-2024-VLDB,LVQ-VLDB-2023,SQ-IR-MM-2012,ExRaBitQ-arxiv-2024,VA-File-plus-CIKM-2000} typically maps \texttt{float32} values to \texttt{int8} and leverages low-precision AVX-512 instructions for efficient computation. 
Binary Quantization (BQ)~\cite{Rabitq-SIGMOD-2024,SRP-STOC-2002} further pushes this idea by mapping each dimension to a single bit, enabling ultra-fast distance computation using bitwise operations (e.g., \texttt{POPCNT}). 
The state-of-the-art RabitQ~\cite{Rabitq-SIGMOD-2024,ExRaBitQ-arxiv-2024} improves accuracy by introducing an unbiased estimator and error bounds for efficient re-ranking.
However, all dimension-wise encoding methods suffer from a \textit{fixed-ratio constraint}: the code length is rigidly tied to the original dimensionality (e.g., a 960-dimensional vector produces a 960-bit binary code). 
This inflexibility prevents users from tuning the compression ratio to meet memory budgets or from aligning code lengths with hardware register widths for optimal SIMD performance.


\section{Problem Analysis and Observation}
This section reiterates the limitations of existing methods and then presents our key observation that motivates the design of $\MRQ$.

\subsection{Problem Analysis}
Existing dimension-wise quantization methods cannot achieve flexible compression ratios because they must quantize \emph{every} dimension of the original vectors. 
A straightforward idea to improve the compression ratio is to truncate or randomly drop some dimensions of the original vectors. 
For example, consider an original vector in \texttt{float32} format: under binary quantization, achieving a $128\times$ compression ratio would require discarding roughly $75\%$ of its dimensions. 
However, such naive truncation introduces severe bias in distance estimation, which in turn leads to significant degradation in AKNN search accuracy.
This raises two natural questions: (1) can we minimize the information loss caused by discarding residual dimensions, and (2) can we still extract useful information from the discarded residual dimensions to further improve accuracy?

\begin{figure}[!t]
\centering\vgap
\begin{small}
\subfloat[DEEP]{
\includegraphics[width=0.495\columnwidth]{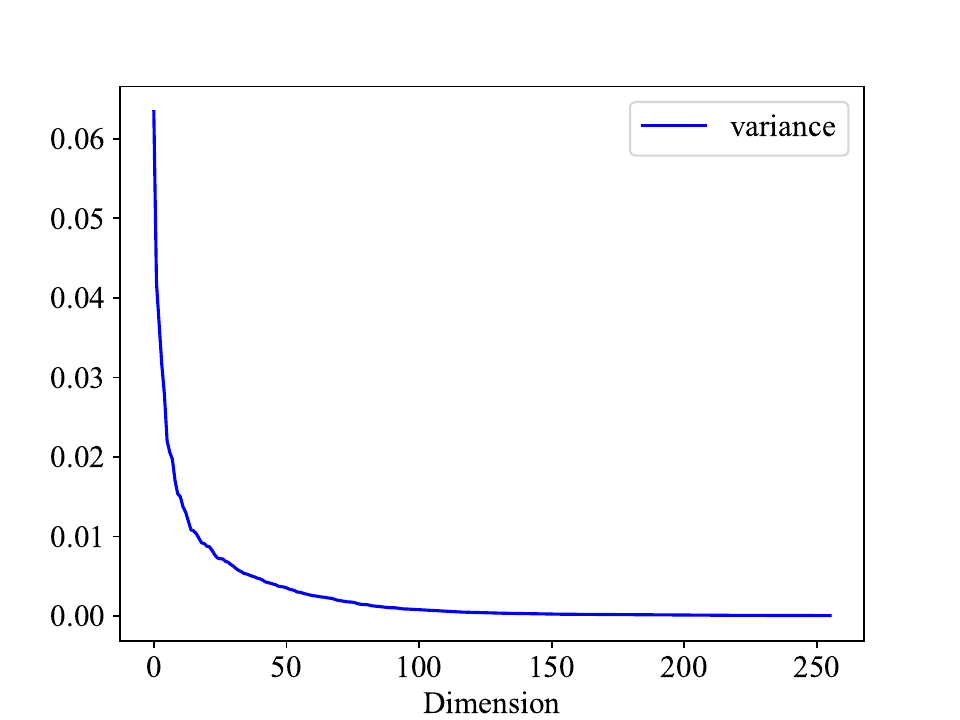}
}
\subfloat[OpenAI-1536]{
\includegraphics[width=0.495\columnwidth]{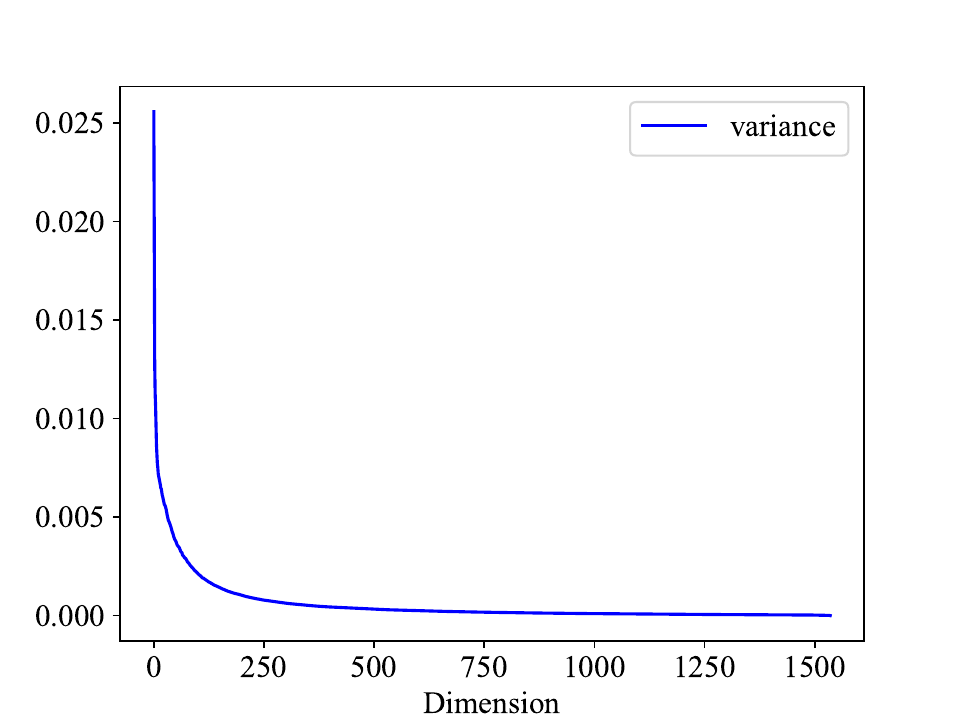}
}
\vgap\caption{Variance Distribution of Vectors After PCA}\label{fig:vec_var_3}
\end{small}\vspace{-1em}
\end{figure}

\subsection{Observation}
To benefit from the significant space reduction offered by quantization methods while overcoming the limitation of a fixed compression ratio, we present the following observation that motivates the development of our new approach.

\stitle{Variance Distribution Analysis.}
We begin our analysis with vector data held in vector databases. 
These data typically comprise high-dimensional embeddings---often exceeding 1,000 dimensions---derived from neural networks for audio, image, and text representations. 
After applying PCA, we observe that the \textbf{variance distribution of the vector data exhibits a long-tailed pattern.}

\begin{exmp}
    We present the observation in Fig.~\ref{fig:vec_var_3}.
    After applying PCA to the text embeddings generated by the OpenAI-1536 model, we observe that the first 512 dimensions capture nearly 90\% of the total variance, while the remaining 1000 dimensions account for only 10\%.
    A similar trend is observed in other data: for image embeddings from the DEEP dataset, only 128 PCA dimensions are sufficient to retain 90\% of all the variance.
\end{exmp}

From an information-theoretic perspective, higher variance in a dimension indicates greater information content. 
Consequently, only a small number of dimensions are required to preserve most of the information in the original vector. 
This suggests that treating all dimensions equally during quantization---such as by randomly dropping or truncating them---is suboptimal compared to using PCA, which more effectively reduces information loss.
Recent studies~\cite{BSA-DDC-Mingyu-ICDE-2025,Liwei-arxiv-DADE} have analyzed PCA also minimizes the variance of the residual dimension, and lower variance in the residual dimensions leads to a smaller mean squared error (MSE) when projecting to a lower-dimensional space. 
Therefore, given their limited information content and low contribution to reconstruction error, these residual dimensions can be safely discarded during quantization. 
Beyond the vector quantization, we observe that the residual part still carries useful information for improving AKNN search accuracy. 
In particular, the norm of the residual can be integrated with the projection part to obtain a more accurate distance estimate, while the variance of the residual provides error bounds for distance approximation. 
In summary, focusing on the high-variance components after PCA not only 
(1) offers a flexible way to discard dimensions while minimizing information loss, but also 
(2) enables the use of residual statistics for further refinement.

\section{Our Quantization Method}
Based on the above observation, we propose a novel quantization method, $\MRQ$ (Minimized Residual Quantization).
In Section~\ref{sub:dec}, we present our vector decomposition strategy, which enables user-controllable compression ratios.
In Section~\ref{sub:app}, we introduce a multi-stage distance computation scheme to ensure high AKNN accuracy.

\begin{figure}[!t]
    \centering\vgap
    \includegraphics[width=0.88\linewidth]{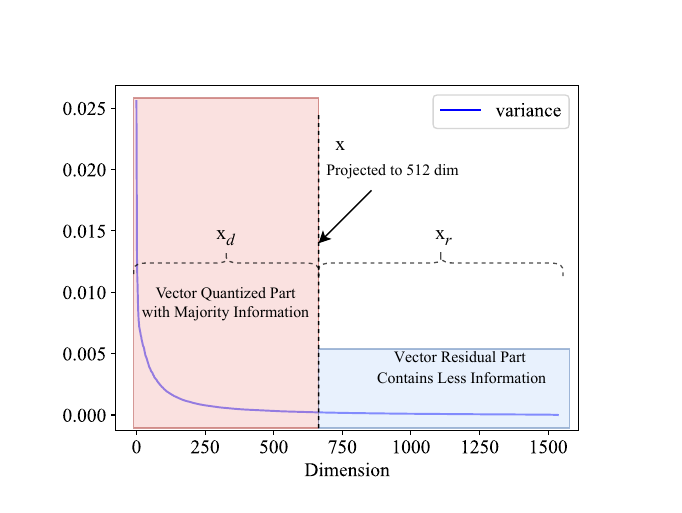}
    \vgap\caption{The Example of Vector Split and Quantization}\vgap
    \label{fig:proj-exmp}
\end{figure}

\subsection{Vector Decomposition}\label{sub:dec}
The observation of variance distribution reveals that after applying PCA to vectors in a dataset, only a small number of dimensions retain most of the information. 
This motivates us to treat the high-variance dimensions differently by \textit{applying quantization only to the projected vectors in this informative subspace}. 
Fig.~\ref{fig:proj-exmp} describes this insight.
Importantly, the number of retained dimensions is fully user-controlled, enabling a flexible compression ratio.

Specifically, consider a vector dataset $S$ consisting of $N$ points in a $D$-dimensional Euclidean space $\mathbb{R}^{D}$.
Each vector $\xx \in S$ can be projected into a lower-dimensional space of dimension $d$ (where $d < D$) using an orthogonal matrix $\mathbf{R}$, i.e., $\xx_d = \mathbf{R}\xx$.
The exact squared Euclidean distance $dis$ between a database vector $\xx$ and a query vector $\qq$ can then be expressed as:
\begin{equation}
    dis = ||\xx -\qq||^2 = ||\xx_d -\qq_d||^2 + ||\xx_r - \qq_r||^2
\end{equation}
where $\xx_r$ is the residual part $(r+d=D)$.
We then apply a quantization method (with state-of-the-art method RabitQ as the example) to the projected vector $\xx_d$.
Specifically, similar to RabitQ, we first normalize and center the vectors after projection.
Let $\cc$ denote the centroid of the projected data $\xx_d$. The normalized projected vector is defined as $\xx_b \coloneq \frac{\xx_d - \cc}{\|\xx_d - \cc\|}$, and the query vector is similarly normalized as $\qq_b \coloneq \frac{\qq_d - \cc}{\|\qq_d - \cc\|}$.
After normalization, the projected distance $\|\xx_d - \qq_d\|^2$ can be expressed in terms of the inner product and vector norms, as formally defined below.

\begin{equation}
\begin{aligned}
    ||\xx_d-\qq_d||^2 & = ||\xx_d -\cc||^2 + ||\qq_d-\cc||^2 \\
                      & -2 \cdot ||\xx_d -\cc|| \cdot ||\qq_d -\cc||\cdot\Braket{\xx_b, \qq_b}     
\end{aligned}
\end{equation}

Note that the inner-product term $\Braket{\xx_d, \qq_d}$ can be approximated by quantizing $\Braket{\xx_b, \qq_b}$ using methods such as RabitQ, yielding a $d$-bit binary representation.
By precomputing the norms and other components in Equation~2, the distance within the projected subspace can be efficiently calculated.
For the residual component $||\xx_r - \qq_r||^2$, it can be decomposed as $||\xx_r||^2 + ||\qq_r||^2 - 2 \cdot \Braket{\xx_r, \qq_r}$. Incorporating this into Equation~2 yields the final formulation for computing the overall distance.

\begin{equation}
\begin{aligned}
    ||\xx-\qq||^2 & = ||\xx_d -\cc||^2 + ||\qq_d-\cc||^2 + ||\xx_r||^2 + ||\qq_r||^2 \\
                & -2 \cdot ||\xx_d -\cc|| \cdot ||\qq_d -\cc||\cdot\Braket{\xx_b, \qq_b} -2 \cdot \Braket{\xx_r, \qq_r}
\end{aligned}
\end{equation}

\begin{figure}[!t]
    \centering
    \includegraphics[width=0.85\linewidth]{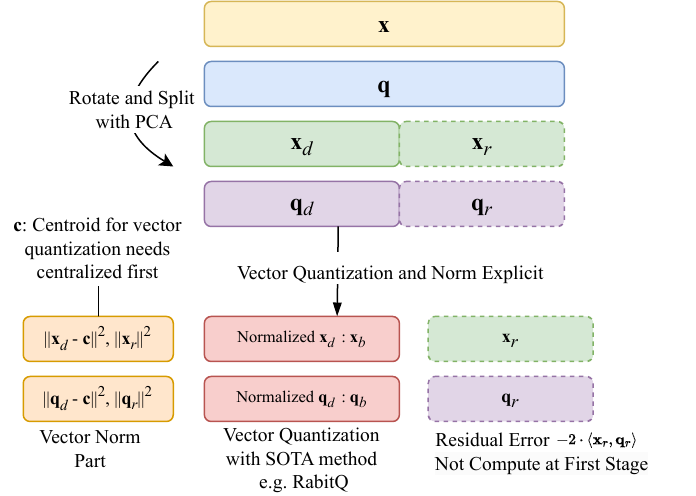}\vspace{-1em}
    \caption{The Example of Vector Decomposition}
    \label{fig:Vec-Deco}
\end{figure}

\begin{example}
Fig.~\ref{fig:Vec-Deco} illustrates an example of Equation~3. 
We first apply PCA to transform the original vectors $\xx$ and $\qq$, yielding their projected components $\xx_d$, $\qq_d$ and residual components $\xx_r$, $\qq_r$. 
The projected components are then normalized and quantized to compute the inner product $\langle \xx_b, \qq_b \rangle$, while the residual components can be precomputed to account for the residual distance term.
\end{example}

\stitle{Remark.}
Equation~3 shows that we only need to quantize the projected vectors $\xx_d$ and $\qq_d$ into $\xx_b$ and $\qq_b$, while leaving the residual vectors $\xx_r$ and $\qq_r$ untouched.
This allows the quantization process to be flexible, as the projection dimension $d$ is user-controllable.


\subsection{Multi-Stage Distance Computation}\label{sub:app}
After vector decomposition, we focus on computing distances for AKNN search.
Given the quantized distance in Equation~2 and the overall distance formulation in Equation~3, we now elaborate on the multi-stage distance computation (as illustrated in Fig.~1).

\stitle{Overview.}
Specifically, we define three components of the distance in Equation~3: (1) The norm component: $C_1 \coloneq ||\xx_d - \cc||^2 + ||\qq_d - \cc||^2 + ||\xx_r||^2 + ||\qq_r||^2$; (2) The quantized component: $C_2 \coloneq -2 \cdot ||\xx_d - \cc|| \cdot ||\qq_d - \cc|| \cdot \Braket{\xx_b, \qq_b}$; and (3) The residual component: $C_3 \coloneq -2 \cdot \Braket{\xx_r, \qq_r}$.
From an efficiency perspective, the norms of the base vector, i.e., $||\xx_d - \cc||$ and $||\xx_r||$, can be precomputed and stored. Consequently, computing $C_1$ at query time only requires calculating $||\qq_d - \cc||$ and $||\qq_r||$, which are computed once per query. With all norms precomputed, the remaining computation reduces to evaluating the inner products $\Braket{\xx_b, \qq_b}$ and $\Braket{\xx_r, \qq_r}$.

As shown in Fig.~1, the multi-stage computation proceeds as follows. 
Stage~(1) estimates the distance using the quantized inner product $\Braket{\xx_b, \qq_b}$ (details provided below). 
This approximation supports efficient filtering: if the estimated distance is already too large, the candidate is discarded. 
Otherwise, we proceed to Stage~(2), where we re-rank candidates using the original projected inner product $\Braket{\xx_d, \qq_d}$, which can be computed directly. 
If even higher precision is required, Stage~(3) computes $\Braket{\xx_r, \qq_r}$ to obtain the exact distance.
In what follows, we focus on Stage~(1)—distance estimation via the quantized inner product $\Braket{\xx_b, \qq_b}$ and the use of error bounds for re-ranking, as Stages~(2) and (3) follow the same reasoning.

\stitle{Distance Estimation.}
The next question is how to estimate the quantized inner product $\Braket{\xx_b, \qq_b}$ for distance approximation.
Note that $\Braket{\xx_b, \qq_b}$ is the inner product between two normalized vectors, where $\qq_b \coloneq \frac{\qq_d - \cc}{|\qq_d - \cc|}$ and $\xx_b \coloneq \frac{\xx_d - \cc}{|\xx_d - \cc|}$.
To estimate this inner product, we follow the design of RabitQ to obtain
the quantized representation of an arbitrary vector $\xx$. 
Specifically, RabitQ implicitly defines a codebook $\mathcal{C} \coloneq \left\{-\tfrac{1}{\sqrt{D}}, +\tfrac{1}{\sqrt{D}}\right\}^{D}$,
and seeks the closest vector $Pu$ to $\xx$, where $P$ is a random orthogonal matrix and $u \in \mathcal{C}$. The vector $\xx$ is then encoded using the
\emph{signs} (positive or negative) of the entries in $u$, yielding the
dimension-wise quantized code $\bar{\xx}_b$ of $\xx_b$.
RabitQ provides an efficient procedure to obtain these signs \emph{without explicitly constructing} the vector $u$, which makes the encoding highly efficient. 
The inner product $\langle \xx_b, \qq_b \rangle$ is then
approximated using the following unbiased estimator:
$\frac{\langle \bar{\xx}_b,\, \qq_b\rangle}
     {\langle \bar{\xx}_b,\, \xx_b\rangle}$.

Using this estimator for $\Braket{\xx_b, \qq_b}$, the remaining term in the distance computation is $\Braket{\xx_r, \qq_r}$. Both empirical evidence and theoretical analysis indicate that the residual component contributes minimally to the total distance. Therefore, we can omit the residual inner product and obtain the final approximate distance:

\begin{equation}\label{eq:app-dist}
\begin{aligned}
    dis^{\prime} & = ||\xx_d -\cc||^2 + ||\qq_d-\cc||^2 + ||\xx_r||^2 + ||\qq_r||^2 \\
                & -2 \cdot ||\xx_d -\cc|| \cdot ||\qq_d -\cc|| \cdot \Braket{\bar{\xx}_b,\qq_b} /\Braket{\bar{\xx}_b,\xx_b}
\end{aligned}
\end{equation}

A key advantage of $\MRQ$ is that the computation of the inner product 
$\langle \bar{\mathbf{x}}_b, \mathbf{q}_b \rangle$ is highly amenable to hardware acceleration. 
Since the projection dimension $d$ can be chosen to match SIMD vector widths (e.g., 128, 256, or 512 bits), 
we can ensure optimal instruction throughput. 
The binary codes $\bar{\mathbf{x}}_b$ can then be processed efficiently using SIMD instructions 
(e.g., \texttt{POPCNT} on AVX2/AVX-512 registers), which are subsequently mapped to inner-product values. 
This stands in contrast to methods whose code length is tied to the original vector dimensionality, 
which may not align with SIMD widths and thus result in suboptimal performance.

\vspace{-0.3em}

\stitle{Error Bounds for Re-Ranking.}
Note that for the re-ranking step introduced in Section~\ref{sub:re-rank}, we estimate a lower bound on the distance between a data point $\xx$ and the query $\qq$, and check whether this bound exceeds the current top-$K$ threshold (i.e., the distance to the $K$-th nearest neighbor found so far). If so, the candidate can be safely pruned. Otherwise, a more accurate distance computation is required, and we need to proceed to the second and third stages of the distance estimation process (see Fig.~1).

We now analyze the error in approximate distance computation, defined as the gap between the true distance $dis$ and the estimated distance $dis^{\prime}$. This error arises from two sources: (1) quantization of the inner product $\Braket{\xx_b, \qq_b}$ introduces quantization error $\epsilon_q$, and (2) omission of the residual dimensions leads to residual error $\epsilon_r$. 
Fortunately, due to the decorrelation introduced by PCA-based rotation, each dimension becomes approximately independent. 
This allows us to analyze $\epsilon_q$ and $\epsilon_r$ separately.
\sstitle{(1) $\epsilon_q$ Part.}
The confidence interval of the estimator $\frac{\Braket{\bar{\xx}, \qq}}{\Braket{\bar{\xx}, \xx}}$ is given by~\cite{Rabitq-SIGMOD-2024}:
\begin{equation}\label{eq:rabit-error}
    \mathbb{P}\left\{\left|\frac{\langle\bar{\xx},\qq\rangle}{\langle\bar{\xx},\xx\rangle}-\langle\xx,\qq\rangle\right|>\sqrt{\frac{1-\left\langle\bar{\xx},\xx\right\rangle^2}{\left\langle\bar{\xx},\xx\right\rangle^2}}\cdot\frac{\epsilon_0}{\sqrt{D-1}}\right\}\leq2e^{-c_0\epsilon_0^2}
\end{equation}
where $\epsilon_0$ is a tunable parameter controlling the failure probability, and $c_0$ is a constant~\cite{Rabitq-SIGMOD-2024}. 
\sstitle{(2) $\epsilon_r$ Part.}
For the residual inner product $\Braket{\xx_r, \qq_r}$, PCA ensures minimal variance in the residual dimensions. A natural approach is to use this variance to bound the error. Let $\sigma_i$ denote the standard deviation of the $i$-th dimension of $\xx$, then the variance of $\Braket{\xx_r, \qq_r}$ can be expressed as:

\begin{equation}\label{eq:var-compute}
    \sigma^2=Var(\Braket{\xx_r,\qq_r}) = \sum_{i=d+1}^{i\leq D}\qq_i^2\sigma_i^2    
\end{equation}
Assuming the data is centered (i.e., $\xx$ has zero mean), we can then apply Chebyshev’s inequality to bound this residual error, that is, $\mathbb{P}(|\Braket{\xx_r, \qq_r}| \ge m\sigma) \le \frac{1}{m^2}$.
Since the quantization and residual errors are independent, their bounds can be combined additively to provide an overall error bound for $\MRQ$ in distance computation.

\section{Integrate with Existing Method}\label{sec:IVF-MRQ}
Our quantization method $\MRQ$ provides an effective way to compute distances, but it is always integrated with existing AKNN search methods for practical deployment. 
As an example, we implement and integrate $\MRQ$ with IVF-based methods, owing to their relatively simple implementation and compact index size.

\subsection{Preprocessing Process}
We first examine the implementation of $\MRQ$ and its integration with the IVF-based indexing structure.
Note that IVF partitions the dataset into disjoint clusters, each represented by a centroid that serves as the index entry.
The overall preprocessing procedure is summarized in Algorithm~\ref{algo:MRQ-Index}.

\stitle{Algorithm.}
We first apply PCA to project the original data into a lower-dimensional space and obtain the variance of the residual dimensions via eigenvalue decomposition (Lines~1–2).
Next, we rotate the original dataset $S$ and compute the norm of the residual vectors (Lines~3–4).
We then construct the IVF index using the projected vectors (Line~5).
The projected vectors are subsequently quantized using a method similar to RabitQ, where the IVF centroids can also serve to centralize the projected data (Lines~7–8).
Note that we use PCA-rotated vectors as base vectors, as Euclidean distances are preserved under the same transformation.
With this setup, all information needed for distance computation is ready for efficient use at query time.

\begin{algorithm}[!t]
	\caption{$\MRQ$ Preprocessing}
	\label{algo:MRQ-Index}
	\begin{footnotesize}
	\KwIn{The database raw vector set $S$ and quantization bit $d$}
	\KwOut{The IVF index $\mathcal{I}$; The PCA matrix $P_{p}$; The random matrix $P_{r}$; The quantization code; The pre-compute results of $||\xx_d -\cc||$, $||\xx_r||$ and $\Braket{\bar{\xx},\xx}$; The residual variance $\sigma_i$}
    Train PCA matrix $P_{p}$ with database vector set $S$\;
    Derive residual variance $\sigma_i$ from PCA training process\;
    Rotate $\xx \in S$ with $P_{p}$ by $\xx_p = P_{p}\xx$\;
    Compute the residual vector norm $||\xx_r||$ from $\xx_p$\;
    Take the first $d$-dimension of $\xx_p$ and get $\xx_d$\;
    Train IVF centroids $\cc_d$ and index $\xx_d$ to get $\mathcal{I}$\;
    Normalized $\xx_d$ and quantized to $\bar{\xx}_d$ with random matrix $P_r$\;
    Compute $||\xx_d -\cc_d||$ and $\Braket{\bar{\xx},\xx}$\;
\end{footnotesize}
\end{algorithm}

\subsection{Query Process}
After preprocessing, we discuss how to conduct the AKNN search under in-memory settings.
The query process is detailed in Algorithm~\ref{algo:MRQ-Query}. 
The parameter $\epsilon_0$ controls the confidence interval in Equation~(\ref{eq:rabit-error}), while $m$ determines the standard deviation bound based on Chebyshev’s inequality. 
The parameter $N^{probe}$ specifies the number of scanned clusters, balancing efficiency and accuracy.

\begin{algorithm}[!t]
	\caption{$\MRQ$ Query}
	\label{algo:MRQ-Query}
	\begin{footnotesize}
	\KwIn{The query vector $\qq$; Index $\mathcal{I}$; $N^{nprobe}$; Quantized vector $\bar{\xx}_d$; Matrix $P_p,P_r$; PCA data $\xx_p$; Error bound parameter $\epsilon_0$ and $m$; The pre-compute results of $||\xx_d -\cc||$, $||\xx_r||$ and $\Braket{\bar{\xx},\xx}$; The residual variance $\sigma_i$}
	\KwOut{The K nearest neighbor of $\qq$}
    Rotate query vector with PCA matrix $\qq_p = P_p\qq$\;
    Take first $d$ dimension of $\qq_p$ as $\qq_d$ and rest as $\qq_r$\;
    Compute residual variance $\sigma$ by Equation~(\ref{eq:var-compute}) with $\qq_r$ and $\sigma_i$\;
    Normalized and random rotate $\qq_d$ by $P_r$ to obtain $\qq_d^{\prime}$\;
    Quantized query $\qq_d^{\prime}$ to $\bar{\qq}$\;
    Result $Q \gets \emptyset$\;
    \For{\textbf{each} $ID\in \mathcal{I}$ \textbf{in} top $N^{probe}$ centroid to $\qq_d$}{
        Compute the approximate distance $dis^{\prime}$ based on Equation~\ref{eq:app-dist}\tcp*{$\MRQ$}
        $\tau \gets$ threshold of result queue $Q$\;
        $\epsilon_b \gets$ RabitQ error with $\epsilon_0$\tcp*{Quantized Error}
        $\epsilon_r \gets m \cdot \sigma$\tcp*{Residual Error}
        \If{$dis^{\prime} - \epsilon_b -\epsilon_r < \tau$}{
            Compute the project distance $dis_o^{\prime}$ based on Equation~\ref{eq:proj-dist}\tcp*{$\MRQ^{+}$}
            \If{$dis_o^{\prime}-\epsilon_r < \tau$}{
                compute $dis=||\xx_p,\qq_p||^2$\;
                update queue $Q$ with $dis$\;   
            }
        }
    }
    \textbf{return} Q\;
\end{footnotesize}
\end{algorithm}

\stitle{Query Algorithm.} 
The query process starts with applying PCA rotation to the query vector $\qq$ (Line~1), followed by computing the residual variance for distance correction (Lines~2–3).
The projected query $\qq_d$ is then quantized to accelerate distance computation (Lines~4–5).
The IVF index ranks cluster centroids by distance and scans vectors in the top $N^{probe}$ clusters to identify the $K$ nearest neighbors of $\qq$, maintaining the top-$K$ candidates in a result queue (heap) $Q$ (Lines~6–7).
For each candidate vector, we first estimate the approximate distance using the quantized representation and precomputed values based on Equation~4 (Line~8).
We then check whether the estimated distance falls below the current threshold with high probability (Lines~9–12).
If it does, we compute the exact distance and update the result queue accordingly (Lines~14–15).

\stitle{Optimizations.}\label{subsec:opt}
In Fig.1, we propose a multi-stage distance computation scheme.
Line~12 of Algorithm~\ref{algo:MRQ-Query} performs stage~(1), while Line~14 performs stage~(3).
We further propose an optimization to include stage~(2): instead of using quantized distance $\Braket{\xx_b, \qq_b}$, we directly use projected distance $\Braket{\xx_d, \qq_d}$ to derive
\begin{equation}\label{eq:proj-dist}
    dis_o^{\prime} = ||\xx||^2 + ||\qq||^2 - 2 \cdot \Braket{\xx_d, \qq_d}.
\end{equation}
Using the residual error bound for re-ranking, we derive the pruning condition $dis_o^{\prime} - m \cdot \sigma < \tau$ (Line~13) to skip unnecessary exact computations.
We also optimize the memory layout following~\cite{ADSampling:journals/sigmod/GaoL23} to enhance cache efficiency.

\subsection{Discussion}\label{subsec:discussion}

\stitle{Parameter Selection.}
A key advantage of $\MRQ$ is the flexibility to choose the projection dimension $d$, enabling it to adapt to different deployment needs.
(1) \textbf{Memory-constrained setting.}
When memory is limited, $d$ can simply be set to the largest value that fits the space budget.
(2) \textbf{Efficiency-oriented setting.}
Recall that $d$ directly affects the multi-stage distance computation, an operation that relies heavily on hardware acceleration (e.g., SIMD). 
Thus, $d$ should be aligned to SIMD-friendly block sizes (e.g., multiples of 64), but not so large as to incur excessive memory usage.
While one could search over $d$ with a step size of 64 to locate a good trade-off, repeatedly rebuilding the index is expensive.
To avoid this overhead, we adopt an empirical rule: choose $d$ as the smallest power of two, $2^{i^\star}$, that (i) is at least $128$ to match SIMD widths, and (ii) captures at least $80\%$ of the total variance:
\begin{equation}\label{eq:para-set}
i^\star \;=\;
\min\left\{\,
i \in \mathbb{N}
:\;
2^{i} \ge 128,\;
\frac{\sum_{j=1}^{2^{i}} \sigma_j}{\sum_{j=1}^{d} \sigma_j}
\ge 0.8
\right\}.
\end{equation}
This rule preserves most information while ensuring small space overhead and, as shown in $\S$~\ref{sec:exp-res}, delivers near-optimal performance.

\stitle{Distributed Deployment.}
When the vector data is extremely large, a natural solution is to distribute the data across a cluster. 
Our proposed method $\MRQ$ readily extends to this setting. 
In particular, when combined with IVF-based methods, $\MRQ$ can be integrated into existing distributed IVF systems such as~\cite{Chameleon-VLDB-2025-SDS,Distribute-Graph-Vector-Index-arxiv-2025-james-cheng,VStream-VLDB-2025,Harmony-SIGMOD-2025}. 
Because $\MRQ$ introduces no additional dependencies on operations such as indexing and query procedures in Section~5, it can be implemented in a parallel or fully distributed manner.
One subtle issue is the use of PCA to obtain global statistics. This can be addressed either by employing distributed PCA~\cite{Distribute-PCA-NIPS-2014} or by sampling a small subset of vectors on a single machine. 
We plan to further explore distributed AKNN search under this framework.


\begin{table}[t!] 
\centering 
\caption{Dataset Statistics}
\label{tab:dataset_details} 
\begin{footnotesize}
\begin{tabular}{c|c c c c} 
\toprule
\textbf{Dataset}& \textbf{Dimension} & \textbf{Size} & \textbf{Query Size} & \textbf{Type} \\ 
\midrule
MSONG & 420 & 992.272 & 200 & Audio \\
GIST & 960 & 1,000,000 & 1000 & Image\\ 
DEEP & 256 & 1,000,000 & 1000 & Image\\ 
TINY & 384 & 5,000,000 & 1000 & Image\\
WORD2VEC & 300 & 1,000,000 & 1000 & Text\\
MSMARCO10M & 1024 & 10,000,000 & 1000 & Text\\
OpenAI-1536& 1536 & 999,000 & 1000 & Text\\
OpenAI-3072 & 3072 & 999,000 & 1000 & Text\\
\bottomrule
\end{tabular} 
\end{footnotesize}
\end{table}

\section{Experiment}\label{sec:exp}

\subsection{Experiment Settings}\label{sec:exp-set}
\stitle{Datasets.}
We evaluate on $8$ public datasets, including widely used benchmarks (MSONG, GIST, DEEP, TINY)\footnote{\url{https://www.cse.cuhk.edu.hk/systems/hash/gqr/datasets.html}} and datasets derived from recent embedding models: MSMARCO10M (10M slice)\footnote{\url{https://huggingface.co/datasets/Cohere/msmarco-v2.1-embed-english-v3}}, OpenAI-1536, and OpenAI-3072\footnote{\url{https://huggingface.co/datasets/Qdrant/dbpedia-entities-openai3-text-embedding-3-large-3072-1M}}.
Table~\ref{tab:dataset_details} summarizes key statistics, including dataset size, query count, dimensionality, and data type (audio, image, or text).
For datasets lacking separate query vectors, we randomly sample 1,000 base vectors as queries and remove them from the base set. All data are stored in float32 format.

\stitle{Evaluation Metrics.}
We evaluate the accuracy using recall@k (see \S~\ref{sec:preliminary}) for AKNN search, with $k$ set to 20. 
For efficiency, we measure queries per second (QPS),
which is the number of queries processed per second.
All metrics are averaged over the entire query set.

\stitle{Algorithms.}
We integrate our quantization method $\MRQ$ with the IVF method, as described in \S~\ref{sec:IVF-MRQ}.
For comparison, we evaluate both graph-based and IVF-based AKNN methods as baselines.
The evaluated algorithms are as follows:

\begin{itemize}[leftmargin=2\labelsep]
\item $\IVF$-RabitQ: IVF combined with the quantization method RabitQ.
\item $\IVF$-$\MRQ$: IVF integrated with our $\MRQ$ method (Algorithm~2 without Lines~13--14).
\item $\IVF$-$\MRQ^{+}$: $\IVF$-$\MRQ$ enhanced with (i) projection-distance optimization (i.e., using Stage~(2) for multi-stage distance computation in Algorithm~2), and (ii) optimized memory layout.
\item HNSW~\cite{DBLP:journals/pami/HNSWMalkovY20}: the most widely used graph-based ANN method.
\end{itemize}
More baselines and results (such as extend $\MRQ$ with SQ and SQ) are provided in the appendix.

\stitle{Configurations.}
We implement the $\MRQ$ method in C++ using g++ version 11.4.0 with the -Ofast optimization flag and -march=native to enable AVX instructions. The number of IVF centroids is set to 4096, following the recommendation of the Faiss library~\cite{faiss:johnson2019billion}.
For the $\IVF$-RabitQ baseline, the quantization bit length is fixed to the original vector dimensionality, as required by its design. 
For all datasets except \texttt{WORD2VEC} and \texttt{MSONG}, we use the raw dimensionality. 
For \texttt{WORD2VEC} and \texttt{MSONG}, we zero-pad the vectors to dimensions 320 and 448, respectively, to enable more efficient SIMD execution.
The quantization bit lengths for $\MRQ$ are determined entirely using our empirical rule in $\S$~5.3:
128 bits for \texttt{MSONG}, \texttt{DEEP}, \texttt{TINY}, and \texttt{GIST};
256 bits for \texttt{WORD2VEC};
and 512 bits for \texttt{OpenAI-1536}, \texttt{OpenAI-3072}, and \texttt{MSMARCO10M}. Our vector decomposition can also be applied to the quantization method Beyond RabitQ to improve efficiency, which has been experimentally verified in the appendix.
We also compared fundamental quantization methods such as PQ and the more competitive OPQ fast-scan~\cite{faiss:johnson2019billion} in the appendix.
For the baseline \HNSW, we use $\kw{M}=16$ and $\kw{efConstruction}=500$, consistent with the recommended settings and recent benchmarks~\cite{DBLP:journals/pami/HNSWMalkovY20,ann-benchmakrs}.

\input{figures/latency-tradeoff-mem}

\begin{figure}[!t]
\centering
\begin{small}
\subfloat[Index Time]{\vgap
\includegraphics[width=0.5\columnwidth]{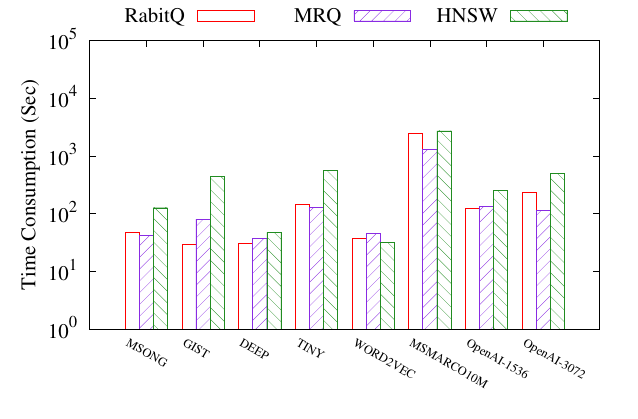}
}
\subfloat[Index Space]{\vgap
\includegraphics[width=0.5\columnwidth]{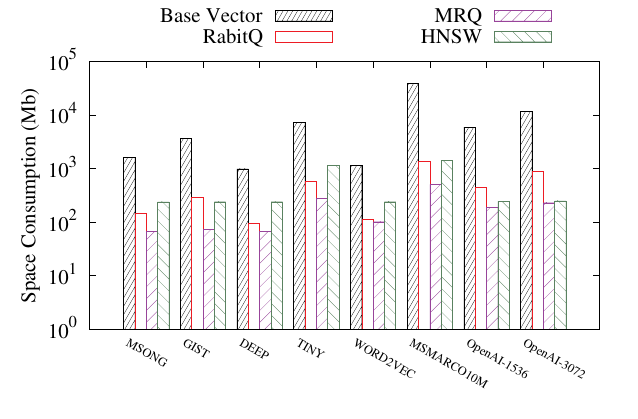}
}
\vgap\caption{The Comparison of the Time Cost and Index Space}\label{fig:Index-time-size}
\end{small}
\end{figure}

\subsection{Experiment Results}\label{sec:exp-res}

\stitle{Exp-1: Test of Performance.}
Fig.~\ref{fig:latency-time-acc-trade} presents the time–accuracy trade-off for all in-memory AKNN search methods, where curves closer to the upper-left corner indicate better performance.
Our key findings are as follows:

\sstitle{(1) Superior efficiency and with fewer quantization bits.}
Our distance computation method $\MRQ$ significantly outperforms RabitQ in efficiency while requiring substantially fewer quantization bits. Specifically: On the GIST dataset, $\MRQ$ achieves a $2\times$ speedup using only 1/7 of the bits; On OpenAI-1536, it delivers a $2.6\times$ improvement with 1/3 of the bits; On OpenAI-3072, it provides a $3.4\times$ gain using just 1/6 of the bits. 
Moreover, IVF-RabitQ$^{+}$ yields almost no performance improvement over IVF-RabitQ, as IVF-RabitQ already prunes most unnecessary distance computations with larger code length, leaving little room for further gains.

\sstitle{(2) Consistent performance advantage over graph-based methods.}
Our method outperforms the widely used graph-based $\HNSW$ in most cases. 
For instance, on GIST, $\MRQ$ is $5.85\times$ faster than $\HNSW$ at high recall levels. While $\HNSW$ performs better on OpenAI-1536 and OpenAI-3072 due to scanning fewer vectors than IVF at equivalent recall, it incurs significantly higher index construction time and memory cost. Moreover, $\HNSW$ performs poorly on WORD2VEC, whereas $\MRQ$ maintains competitive performance across all datasets.

\stitle{Exp-2: Preprocessing Time and Space.}
We report the preprocessing time and space cost (excluding base vectors) for all compared AKNN methods in Fig.~\ref{fig:Index-time-size}.

\sstitle{(1) Preprocessing Time.}
As shown in Fig.~\ref{fig:Index-time-size}~(a), our method achieves significantly faster index construction. In particular, it requires only half the time of RabitQ and one-quarter of $\HNSW$.

\sstitle{(2) Preprocessing Space.}
Fig.~\ref{fig:Index-time-size}~(b) shows that our method also incurs substantially lower space overhead. For example, it consumes over $5\times$ less memory than $\HNSW$ and $2\times$ less than $\IVF$-RabitQ.
Moreover, IVF-RabitQ can exceed the size of HNSW because its quantized codes become large in high-dimensional settings, whereas our method consistently uses substantially less memory.
The advantages in both preprocessing time and size make our method particularly suitable for resource-constrained environments.

\input{figures/indepth-analysis}
\stitle{Exp-3: Understanding Effectiveness.}
To clarify why the proposed method $\MRQ$ is effective for AKNN search, we conduct an in-depth analysis against the state-of-the-art quantization method RabitQ. 
We measure the \emph{prune ratio}, defined as the fraction of pruned distance computations over all exact distance computations, where a larger ratio indicates greater effectiveness because more unnecessary computations are removed.

Fig.~\ref{fig:in-depth-analysis} shows that RabitQ can prune more than $99\%$ of exact distance computations (y-axis: prune rate using original vectors). 
Although RabitQ is strong in pruning, its fixed compression ratio prevents it from operating efficiently with fewer bits. 
Even when we adapt the same distance-optimization and memory-layout techniques used in $\MRQ$ (i.e., to form RabitQ+), the overall speedup remains limited, as also confirmed in Fig.~5.
In contrast, $\MRQ$ uses only $1/8$--$1/3$ as many bits while still pruning $96\%$--$98\%$ of exact distance computations—losing only $1$--$2\%$ compared with RabitQ. 
Moreover, the second stage distance computation in $\MRQ^+$ leverages higher-precision PCA projection distances with nearly $100\%$ prune rate, leading to additional efficiency gains.

\input{figures/parameter-select}
\stitle{Exp-4: Parameter Selection.}
To validate the parameter-selection strategy introduced in $\S$~\ref{subsec:discussion}, we evaluate its effectiveness in real AKNN search scenarios. 
Specifically, we report the search performance obtained from a grid search over the projection dimension $d$ with a step size of 64 (Fig.~\ref{fig:parameter-select}).
From the variance distribution in Fig.~\ref{fig:vec_var_3}, the \texttt{GIST} dataset requires the first 78 projected dimensions to capture 80\% of the total variance, whereas \texttt{OpenAI-1536} requires 354. 
Following our empirical rule—which aligns $d$ to the nearest multiple of 64 for efficient SIMD execution—we set $d = 128$ for \texttt{GIST} and $d = 512$ for \texttt{OpenAI-1536}. 
The efficiency curves in Fig.~\ref{fig:parameter-select} show that these empirically chosen values (solid-line markers) achieve near-optimal search performance across the grid. 
This empirical rule is used throughout all experiments in Fig.~\ref{fig:latency-time-acc-trade}, offering a low-cost yet effective guideline for efficiency-oriented index configuration.

\section{Conclusion}
This paper proposes a novel method, $\MRQ$, for AKNN search.
$\MRQ$ decomposes each vector into quantized and residual components via projection.
The quantized component operates on user-controllable dimensions, enabling arbitrary compression ratios.
The residual component is modeled using lightweight statistical summaries, further reducing memory usage.
By combining both components, $\MRQ$ performs a multi-stage distance computation tailored for efficient and accurate AKNN search.
Extensive experiments show that $\MRQ$ greatly improves search efficiency with negligible index construction time and storage overhead.
Future work will investigate distributed processing techniques for large-scale vector datasets.

\begin{acks}
We are grateful to the anonymous reviewers for their constructive comments. Wentao Li is supported by NSFC (Grant No.62302417).
This work is supported by Advanced Materials-National Science and Technology Major Project (Grant No. 2025ZD0620100), HKUST(GZ)-IEIP-RoP (G01RF000256) and Ant Group.
\end{acks}

\balance
\bibliographystyle{ACM-Reference-Format}
\bibliography{sample}
\clearpage
\input{appendix}

\end{document}

%% file: figures/latency-tradeoff-mem.tex
\begin{figure*}[t!]
\centering
\begin{small}
\begin{tikzpicture}
    \begin{customlegend}[legend columns=5,
        legend entries={$\IVF$-RabitQ,$\IVF$-$\MRQ$,$\IVF$-$\MRQ^{+}$,$\HNSW$},
        legend style={at={(0.45,1.15)},anchor=north,draw=none,font=\scriptsize,column sep=0.1cm}]
    \addlegendimage{line width=0.15mm,color=amaranth,mark=o,mark size=0.5mm}
    \addlegendimage{line width=0.15mm,color=amber,mark=triangle,mark size=0.5mm}
    \addlegendimage{line width=0.15mm,color=violate,mark=square,mark size=0.5mm}
    \addlegendimage{line width=0.15mm,color=forestgreen,mark=pentagon,mark size=0.5mm}
    \end{customlegend}
\end{tikzpicture}
\\[-\lineskip]

\subfloat[GIST]{\vspace{-2mm}
\begin{tikzpicture}[scale=1]
\begin{axis}[
    height=\columnwidth/2.70,
width=\columnwidth/1.90,
xlabel=recall@20,
ylabel=Qpsx100,
label style={font=\scriptsize},
tick label style={font=\scriptsize},
ymajorgrids=true,
xmajorgrids=true,
grid style=dashed,
]
\addplot[line width=0.15mm,color=amaranth,mark=o,mark size=0.5mm]
plot coordinates {
(0.70025, 11.0196)
(0.8272, 9.48403)
(0.8869, 8.38973)
(0.9189, 7.48001)
(0.94025, 6.75347)
(0.9537, 6.28338)
(0.9638, 5.8112)
(0.97095, 5.40643)
(0.9776, 4.99749)
(0.9816, 4.74842)
(0.98475, 4.1899)
(0.98725, 4.18877)
(0.9896, 3.95406)
(0.9912, 3.78924)
(0.99205, 3.62139)
(0.99295, 3.43964)
(0.9935, 3.30898)
(0.9943, 3.17762)
(0.99495, 3.07247)
(0.99535, 2.9571)
};
\addplot[line width=0.15mm,color=amber,mark=triangle,mark size=0.5mm]
plot coordinates {
    ( 0.689, 20.532 )
    ( 0.817, 16.099 )
    ( 0.881, 13.534 )
    ( 0.914, 12.24 )
    ( 0.937, 11.111 )
    ( 0.952, 10.318 )
    ( 0.962, 9.79 )
    ( 0.968, 9.285 )
    ( 0.974, 8.85 )
    ( 0.979, 8.447 )
    ( 0.982, 8.071 )
    ( 0.986, 7.874 )
    ( 0.987, 7.526 )
    ( 0.989, 7.358 )
    ( 0.991, 7.054 )
    ( 0.992, 6.884 )
    ( 0.994, 6.574 )
    ( 0.995, 6.344 )
};
\addplot[line width=0.15mm,color=violate,mark=square,mark size=0.5mm]
plot coordinates {
    ( 0.686, 30.532 )
    ( 0.813, 26.016 )
    ( 0.876, 22.42 )
    ( 0.908, 20.05 )
    ( 0.931, 18.348 )
    ( 0.946, 16.876 )
    ( 0.956, 15.929 )
    ( 0.962, 12.038 )
    ( 0.979, 10.548 )
    ( 0.981, 10.311 )
    ( 0.982, 9.725 )
    ( 0.984, 9.475 )
    ( 0.985, 9.241 )
    ( 0.987, 9.588 )
    ( 0.988, 8.981 )
};
\addplot[line width=0.15mm,color=forestgreen,mark=pentagon,mark size=0.5mm]
plot coordinates {
    ( 0.911, 4.805 )
    ( 0.97, 3.28 )
    ( 0.986, 2.459 )
    ( 0.991, 1.875 )
    ( 0.994, 1.626 )
    ( 0.996, 1.441 )
};
\end{axis}
\end{tikzpicture}\hspace{2mm}
}
\subfloat[MSONG]{\vspace{-2mm}
\begin{tikzpicture}[scale=1]
\begin{axis}[
    height=\columnwidth/2.70,
width=\columnwidth/1.90,
xlabel=recall@20,
ylabel=Qpsx1000,
label style={font=\scriptsize},
tick label style={font=\scriptsize},
ymajorgrids=true,
xmajorgrids=true,
grid style=dashed,
]
\addplot[line width=0.15mm,color=amaranth,mark=o,mark size=0.5mm]
plot coordinates {
    ( 0.752, 2.991 )
    ( 0.886, 2.936 )
    ( 0.933, 2.799 )
    ( 0.957, 2.61 )
    ( 0.97, 2.571 )
    ( 0.978, 2.482 )
    ( 0.983, 2.398 )
    ( 0.987, 2.359 )
    ( 0.989, 2.201 )
    ( 0.991, 2.184 )
    ( 0.993, 2.137 )
    ( 0.994, 2.061 )
    ( 0.996, 1.967 )
    ( 0.997, 1.85 )
    ( 0.998, 1.77 )
};
\addplot[line width=0.15mm,color=amber,mark=triangle,mark size=0.5mm]
plot coordinates {
    ( 0.749, 6.862 )
    ( 0.884, 6.109 )
    ( 0.934, 5.734 )
    ( 0.956, 5.135 )
    ( 0.97, 4.868 )
    ( 0.977, 4.557 )
    ( 0.983, 4.427 )
    ( 0.986, 4.208 )
    ( 0.99, 4.208 )
    ( 0.992, 4.004 )
    ( 0.993, 3.87 )
    ( 0.994, 3.711 )
    ( 0.995, 3.548 )
    ( 0.997, 3.376 )
    ( 0.998, 3.064 )
};
\addplot[line width=0.15mm,color=violate,mark=square,mark size=0.5mm]
plot coordinates {
    ( 0.749, 6.965 )
    ( 0.883, 6.157 )
    ( 0.933, 5.713 )
    ( 0.955, 5.391 )
    ( 0.968, 5.099 )
    ( 0.976, 4.921 )
    ( 0.981, 4.597 )
    ( 0.985, 4.532 )
    ( 0.988, 4.316 )
    ( 0.99, 4.234 )
    ( 0.992, 4.059 )
    ( 0.993, 4.002 )
    ( 0.994, 3.737 )
    ( 0.995, 3.533 )
    ( 0.996, 3.259 )
};
\addplot[line width=0.15mm,color=forestgreen,mark=pentagon,mark size=0.5mm]
plot coordinates {
    ( 0.897, 7.217 )
    ( 0.957, 4.565 )
    ( 0.978, 3.214 )
    ( 0.987, 2.537 )
    ( 0.992, 2.12 )
    ( 0.994, 1.828 )
    ( 0.995, 1.612 )
};
\end{axis}
\end{tikzpicture}\hspace{2mm}
}
\subfloat[OpenAI-1536]{\vspace{-2mm}
\begin{tikzpicture}[scale=1]
\begin{axis}[
    height=\columnwidth/2.70,
width=\columnwidth/1.90,
xlabel=recall@20,
ylabel=Qpsx100,
label style={font=\scriptsize},
tick label style={font=\scriptsize},
ymajorgrids=true,
xmajorgrids=true,
grid style=dashed,
]
\addplot[line width=0.15mm,color=amaranth,mark=o,mark size=0.5mm]
plot coordinates {
    ( 0.863, 8.281 )
    ( 0.908, 7.151 )
    ( 0.93, 6.28 )
    ( 0.945, 5.642 )
    ( 0.953, 5.056 )
    ( 0.961, 4.585 )
    ( 0.966, 3.214 )
    ( 0.97, 3.06 )
    ( 0.973, 3.144 )
    ( 0.975, 2.949 )
    ( 0.977, 2.826 )
    ( 0.979, 2.673 )
    ( 0.981, 2.53 )
    ( 0.983, 2.407 )
    ( 0.984, 2.293 )
    ( 0.986, 2.19 )
    ( 0.987, 2.098 )
    ( 0.988, 2.013 )
    ( 0.989, 1.862 )
};
\addplot[line width=0.15mm,color=amber,mark=triangle,mark size=0.5mm]
plot coordinates {
    ( 0.864, 14.138 )
    ( 0.911, 12.577 )
    ( 0.933, 11.395 )
    ( 0.947, 10.502 )
    ( 0.956, 9.79 )
    ( 0.963, 9.107 )
    ( 0.967, 8.547 )
    ( 0.971, 8.067 )
    ( 0.974, 7.628 )
    ( 0.977, 7.291 )
    ( 0.978, 6.915 )
    ( 0.981, 6.612 )
    ( 0.982, 6.31 )
    ( 0.984, 6.042 )
    ( 0.985, 5.822 )
    ( 0.987, 5.59 )
    ( 0.988, 5.383 )
    ( 0.989, 5.015 )
};
\addplot[line width=0.15mm,color=violate,mark=square,mark size=0.5mm]
plot coordinates {
    ( 0.862, 14.399 )
    ( 0.908, 12.923 )
    ( 0.931, 11.708 )
    ( 0.944, 10.649 )
    ( 0.953, 9.994 )
    ( 0.96, 9.33 )
    ( 0.965, 8.735 )
    ( 0.968, 8.222 )
    ( 0.972, 7.766 )
    ( 0.974, 7.418 )
    ( 0.976, 7.008 )
    ( 0.978, 6.689 )
    ( 0.98, 6.419 )
    ( 0.981, 6.115 )
    ( 0.983, 5.875 )
    ( 0.984, 5.644 )
    ( 0.985, 5.423 )
    ( 0.987, 5.047 )
};
\addplot[line width=0.15mm,color=forestgreen,mark=pentagon,mark size=0.5mm]
plot coordinates {
    ( 0.969, 9.12 )
    ( 0.989, 5.131 )
    ( 0.995, 3.628 )
    ( 0.997, 2.843 )
};
\end{axis}
\end{tikzpicture}\hspace{2mm}
}
\subfloat[OpenAI-3072]{\vspace{-2mm}
\begin{tikzpicture}[scale=1]
\begin{axis}[
    height=\columnwidth/2.70,
width=\columnwidth/1.90,
xlabel=recall@20,
ylabel=Qpsx100,
label style={font=\scriptsize},
tick label style={font=\scriptsize},
ymajorgrids=true,
xmajorgrids=true,
grid style=dashed,
]
\addplot[line width=0.15mm,color=amaranth,mark=o,mark size=0.5mm]
plot coordinates {
    ( 0.85, 2.894 )
    ( 0.893, 2.568 )
    ( 0.919, 2.314 )
    ( 0.935, 2.112 )
    ( 0.944, 1.95 )
    ( 0.952, 1.808 )
    ( 0.957, 1.69 )
    ( 0.963, 1.588 )
    ( 0.966, 1.497 )
    ( 0.969, 1.417 )
    ( 0.971, 1.345 )
    ( 0.973, 1.28 )
    ( 0.976, 1.219 )
    ( 0.977, 1.166 )
    ( 0.979, 1.119 )
    ( 0.98, 1.077 )
    ( 0.982, 1.035 )
    ( 0.983, 0.997 )
    ( 0.985, 0.929 )
};
\addplot[line width=0.15mm,color=amber,mark=triangle,mark size=0.5mm]
plot coordinates {
    ( 0.851, 6.905 )
    ( 0.895, 6.444 )
    ( 0.919, 6.12 )
    ( 0.935, 5.818 )
    ( 0.945, 5.555 )
    ( 0.952, 5.356 )
    ( 0.957, 5.149 )
    ( 0.961, 4.93 )
    ( 0.964, 4.787 )
    ( 0.967, 4.639 )
    ( 0.97, 4.489 )
    ( 0.972, 4.333 )
    ( 0.973, 4.218 )
    ( 0.975, 4.101 )
    ( 0.977, 3.974 )
    ( 0.978, 3.867 )
    ( 0.979, 3.768 )
    ( 0.98, 3.667 )
    ( 0.982, 2.985 )
};
\addplot[line width=0.15mm,color=violate,mark=square,mark size=0.5mm]
plot coordinates {
    ( 0.845, 7.352 )
    ( 0.888, 6.965 )
    ( 0.912, 6.622 )
    ( 0.927, 6.305 )
    ( 0.938, 6.022 )
    ( 0.944, 5.85 )
    ( 0.949, 5.599 )
    ( 0.952, 5.41 )
    ( 0.955, 5.012 )
    ( 0.958, 4.863 )
    ( 0.961, 4.899 )
    ( 0.963, 4.774 )
    ( 0.964, 4.606 )
    ( 0.966, 4.496 )
    ( 0.967, 4.325 )
    ( 0.968, 4.245 )
    ( 0.969, 4.187 )
    ( 0.97, 4.017 )
    ( 0.972, 3.832 )
};
\addplot[line width=0.15mm,color=forestgreen,mark=pentagon,mark size=0.5mm]
plot coordinates {
    ( 0.969, 5.086 )
    ( 0.99, 2.884 )
    ( 0.995, 2.041 )
    ( 0.997, 1.595 )
};
\end{axis}
\end{tikzpicture}\hspace{2mm}
}
\\
\subfloat[DEEP]{\vspace{-2mm}
\begin{tikzpicture}[scale=1]
\begin{axis}[
    height=\columnwidth/2.70,
width=\columnwidth/1.90,
xlabel=recall@20,
ylabel=Qpsx100,
label style={font=\scriptsize},
tick label style={font=\scriptsize},
ymajorgrids=true,
xmajorgrids=true,
grid style=dashed,
]
\addplot[line width=0.15mm,color=amaranth,mark=o,mark size=0.5mm]
plot coordinates {
(0.7881, 47.3608)
(0.8797, 42.1622)
(0.91965, 36.4016)
(0.9421, 34.1142)
(0.9543, 32.3877)
(0.9628, 29.7929)
(0.97065, 28.6572)
(0.9761, 25.8645)
(0.9803, 24.7353)
(0.98295, 23.128)
(0.98515, 22.1476)
(0.98725, 20.9815)
(0.98865, 19.7699)
(0.98985, 19.4647)
(0.991, 18.703)
(0.9919, 18.2366)
(0.99275, 17.8507)
(0.9934, 17.2581)
(0.99355, 13.515)
(0.99385, 10.5818)
};
\addplot[line width=0.15mm,color=amber,mark=triangle,mark size=0.5mm]
plot coordinates {
    ( 0.788, 68.22 )
    ( 0.88, 54.552 )
    ( 0.92, 47.58 )
    ( 0.943, 42.599 )
    ( 0.955, 38.867 )
    ( 0.964, 36.305 )
    ( 0.97, 33.619 )
    ( 0.976, 31.535 )
    ( 0.979, 29.867 )
    ( 0.982, 28.383 )
    ( 0.984, 27.152 )
    ( 0.985, 25.821 )
    ( 0.987, 24.877 )
    ( 0.988, 23.819 )
    ( 0.99, 23.023 )
    ( 0.99, 22.304 )
    ( 0.992, 20.934 )
};
\addplot[line width=0.15mm,color=violate,mark=square,mark size=0.5mm]
plot coordinates {
    ( 0.876, 58.776 )
    ( 0.915, 54.511 )
    ( 0.937, 47.717 )
    ( 0.95, 44.45 )
    ( 0.959, 40.109 )
    ( 0.965, 37.312 )
    ( 0.97, 34.982 )
    ( 0.973, 33.528 )
    ( 0.976, 31.286 )
    ( 0.978, 30.365 )
    ( 0.98, 28.717 )
    ( 0.981, 27.33 )
    ( 0.982, 26.169 )
    ( 0.984, 25.216 )
    ( 0.985, 23.336 )
    ( 0.986, 21.921 )
};
\addplot[line width=0.15mm,color=forestgreen,mark=pentagon,mark size=0.5mm]
plot coordinates {
    ( 0.862, 58.718 )
    ( 0.936, 33.162 )
    ( 0.962, 23.697 )
    ( 0.975, 18.308 )
    ( 0.983, 15.058 )
    ( 0.987, 12.923 )
    ( 0.99, 11.233 )
    ( 0.992, 10.052 )
    ( 0.994, 9.034 )
    ( 0.995, 8.274 )
};
\end{axis}
\end{tikzpicture}\hspace{2mm}
}
\subfloat[WORD2VEC]{\vspace{-2mm}
\begin{tikzpicture}[scale=1]
\begin{axis}[
    height=\columnwidth/2.70,
width=\columnwidth/1.90,
xlabel=recall@20,
ylabel=Qpsx100,
label style={font=\scriptsize},
tick label style={font=\scriptsize},
ymajorgrids=true,
xmajorgrids=true,
grid style=dashed,
]
\addplot[line width=0.15mm,color=amaranth,mark=o,mark size=0.5mm]
plot coordinates {
    ( 0.865, 28.679 )
    ( 0.945, 23.768 )
    ( 0.971, 20.541 )
    ( 0.981, 18.338 )
    ( 0.986, 15.978 )
    ( 0.99, 14.393 )
    ( 0.991, 13.721 )
    ( 0.992, 12.69 )
    ( 0.993, 11.723 )
    ( 0.994, 10.427 )
    ( 0.995, 8.437 )
};
\addplot[line width=0.15mm,color=amber,mark=triangle,mark size=0.5mm]
plot coordinates {
    ( 0.858, 30.879 )
    ( 0.945, 24.133 )
    ( 0.97, 20.56 )
    ( 0.982, 18.236 )
    ( 0.987, 16.539 )
    ( 0.991, 15.323 )
    ( 0.993, 14.086 )
    ( 0.994, 13.026 )
    ( 0.995, 12.178 )
    ( 0.997, 10.65 )
    ( 0.998, 8.948 )
};
\addplot[line width=0.15mm,color=violate,mark=square,mark size=0.5mm]
plot coordinates {
    ( 0.858, 30.477 )
    ( 0.944, 24.613 )
    ( 0.969, 20.983 )
    ( 0.981, 18.549 )
    ( 0.987, 16.776 )
    ( 0.99, 15.407 )
    ( 0.992, 14.182 )
    ( 0.993, 13.26 )
    ( 0.995, 12.395 )
    ( 0.996, 10.903 )
    ( 0.997, 8.645 )
};
\addplot[line width=0.15mm,color=forestgreen,mark=pentagon,mark size=0.5mm]
plot coordinates {
    ( 0.769, 8.132 )
    ( 0.804, 4.379 )
    ( 0.822, 3.348 )
    ( 0.831, 2.565 )
    ( 0.838, 2.093 )
    ( 0.843, 1.78 )
    ( 0.847, 1.552 )
    ( 0.85, 1.377 )
    ( 0.853, 1.239 )
    ( 0.856, 1.126 )
    ( 0.858, 1.035 )
    ( 0.86, 0.957 )
    ( 0.861, 0.889 )
    ( 0.863, 0.833 )
    ( 0.864, 0.783 )
};
\end{axis}
\end{tikzpicture}\hspace{2mm}

}
\subfloat[MSMARCO10M]{\vspace{-2mm}
\begin{tikzpicture}[scale=1]
\begin{axis}[
    height=\columnwidth/2.70,
width=\columnwidth/1.90,
xlabel=recall@20,
ylabel=Qpsx100,
label style={font=\scriptsize},
tick label style={font=\scriptsize},
ymajorgrids=true,
xmajorgrids=true,
grid style=dashed,
]
\addplot[line width=0.15mm,color=amaranth,mark=o,mark size=0.5mm]
plot coordinates {
    ( 0.921, 6.375 )
    ( 0.956, 3.867 )
    ( 0.969, 2.854 )
    ( 0.976, 2.287 )
    ( 0.981, 1.864 )
    ( 0.985, 1.595 )
    ( 0.988, 1.398 )
    ( 0.989, 1.235 )
    ( 0.991, 1.119 )
    ( 0.992, 1.019 )
    ( 0.994, 0.863 )
    ( 0.995, 0.746 )
    ( 0.996, 0.658 )
    ( 0.997, 0.534 )
};
\addplot[line width=0.15mm,color=amber,mark=triangle,mark size=0.5mm]
plot coordinates {
    ( 0.948, 3.914 )
    ( 0.973, 2.741 )
    ( 0.983, 2.314 )
    ( 0.987, 1.987 )
    ( 0.99, 1.718 )
    ( 0.993, 1.543 )
    ( 0.994, 1.41 )
    ( 0.997, 1.105 )
    ( 0.998, 0.968 )
};
\addplot[line width=0.15mm,color=violate,mark=square,mark size=0.5mm]
plot coordinates {
    ( 0.948, 5.433 )
    ( 0.973, 3.965 )
    ( 0.982, 3.163 )
    ( 0.987, 2.557 )
    ( 0.99, 2.236 )
    ( 0.992, 1.962 )
    ( 0.994, 1.719 )
    ( 0.995, 1.458 )
    ( 0.997, 1.242 )
    ( 0.997, 1.089 )
};
\addplot[line width=0.15mm,color=forestgreen,mark=pentagon,mark size=0.5mm]
plot coordinates {
    ( 0.981, 2.709 )
    ( 0.991, 1.492 )
    ( 0.995, 1.039 )
    ( 0.996, 0.798 )
};
\end{axis}
\end{tikzpicture}\hspace{2mm}
}
\subfloat[TINY]{\vspace{-2mm}
\begin{tikzpicture}[scale=1]
\begin{axis}[
    height=\columnwidth/2.70,
width=\columnwidth/1.90,
xlabel=recall@20,
ylabel=Qpsx100,
label style={font=\scriptsize},
tick label style={font=\scriptsize},
ymajorgrids=true,
xmajorgrids=true,
grid style=dashed,
]
\addplot[line width=0.15mm,color=amaranth,mark=o,mark size=0.5mm]
plot coordinates {
    ( 0.773, 10.411 )
    ( 0.871, 9.966 )
    ( 0.913, 8.395 )
    ( 0.936, 7.079 )
    ( 0.951, 5.921 )
    ( 0.961, 5.202 )
    ( 0.97, 4.716 )
    ( 0.976, 4.355 )
    ( 0.98, 3.955 )
    ( 0.983, 3.629 )
    ( 0.986, 3.318 )
    ( 0.988, 3.077 )
    ( 0.99, 2.876 )
    ( 0.992, 2.695 )
    ( 0.993, 2.546 )
    ( 0.994, 2.287 )
    ( 0.995, 2.072 )
};
\addplot[line width=0.15mm,color=amber,mark=triangle,mark size=0.5mm]
plot coordinates {
    ( 0.775, 19.252 )
    ( 0.868, 13.605 )
    ( 0.911, 10.939 )
    ( 0.936, 9.375 )
    ( 0.951, 8.206 )
    ( 0.962, 6.077 )
    ( 0.97, 6.634 )
    ( 0.976, 6.112 )
    ( 0.98, 5.655 )
    ( 0.984, 5.282 )
    ( 0.987, 4.99 )
    ( 0.989, 4.715 )
    ( 0.991, 4.471 )
    ( 0.992, 4.258 )
    ( 0.993, 4.054 )
    ( 0.994, 3.888 )
    ( 0.995, 3.588 )
};
\addplot[line width=0.15mm,color=violate,mark=square,mark size=0.5mm]
plot coordinates {
    ( 0.77, 25.103 )
    ( 0.862, 17.528 )
    ( 0.904, 13.928 )
    ( 0.928, 11.661 )
    ( 0.942, 10.075 )
    ( 0.952, 8.703 )
    ( 0.96, 8.091 )
    ( 0.966, 7.381 )
    ( 0.97, 6.755 )
    ( 0.973, 6.287 )
    ( 0.976, 5.852 )
    ( 0.978, 5.508 )
    ( 0.98, 5.141 )
    ( 0.982, 4.69 )
    ( 0.983, 4.463 )
    ( 0.984, 4.074 )
};
\addplot[line width=0.15mm,color=forestgreen,mark=pentagon,mark size=0.5mm]
plot coordinates {
    ( 0.94, 4.574 )
    ( 0.971, 2.443 )
    ( 0.982, 1.703 )
    ( 0.986, 1.317 )
    ( 0.989, 1.061 )
    ( 0.992, 0.834 )
    ( 0.993, 0.564 )
    ( 0.994, 0.63 )
    ( 0.995, 0.587 )
};
\end{axis}
\end{tikzpicture}\hspace{2mm}
}

\vgap\caption{The Performance Test Among Various Method}\vgap\label{fig:latency-time-acc-trade}
\end{small}
\end{figure*}

%% file: figures/indepth-analysis.tex
\begin{figure}[!t]
\centering
\begin{small}
\begin{tikzpicture}
    \begin{customlegend}[legend columns=3,
        legend entries={RabitQ, $\MRQ$, PCA-$\MRQ^+$},
        legend style={at={(0.45,1.15)},anchor=north,draw=none,font=\scriptsize,column sep=0.1cm}]
    \addlegendimage{line width=0.15mm,color=amaranth,mark=o,mark size=0.5mm}
    \addlegendimage{line width=0.15mm,color=amber,mark=triangle,mark size=0.5mm}
    \addlegendimage{line width=0.15mm,color=violate,mark=square,mark size=0.5mm}
    \end{customlegend}
\end{tikzpicture}
\\[-\lineskip]

\subfloat[GIST]{\vgap
\begin{tikzpicture}[scale=1]
\begin{axis}[
    height=\columnwidth/2.70,
    width=\columnwidth/2.0,
    xlabel=recall@20 (\%),
    ylabel=Prune-Rate (\%),
    label style={font=\scriptsize},
    tick label style={font=\scriptsize},
    xmin=90, xmax=100,
]
\addplot[width=0.15mm,color=amaranth,mark=o,mark size=0.5mm]
coordinates {
    (70.05, 97.41186)
    (82.73, 98.57504)
    (88.705, 99.00325)
    (91.905, 99.22976)
    (94.04, 99.37120)
    (95.385, 99.46799)
    (96.385, 99.53847)
    (97.10, 99.59231)
    (97.755, 99.63466)
    (98.155, 99.66865)
    (98.47, 99.69682)
    (98.72, 99.72042)
    (98.955, 99.74065)
    (99.115, 99.75803)
    (99.20, 99.77323)
    (99.29, 99.78657)
    (99.345, 99.79845)
    (99.425, 99.80898)
    (99.49, 99.81846)
    (99.53, 99.82702)
};
\addplot[width=0.15mm,color=amber,mark=triangle,mark size=0.5mm]
coordinates {
    (68.885, 90.05052)
    (81.815, 93.49540)
    (88.035, 94.98361)
    (91.32, 95.86387)
    (93.75, 96.45359)
    (95.225, 96.87947)
    (96.22, 97.20788)
    (96.78, 97.46819)
    (97.485, 97.67948)
    (97.92, 97.85633)
    (98.265, 98.00596)
    (98.565, 98.13517)
    (98.775, 98.24747)
    (98.96, 98.34670)
    (99.11, 98.43475)
    (99.225, 98.51310)
    (99.315, 98.58449)
    (99.39, 98.64884)
    (99.46, 98.70720)
    (99.50, 98.76083)
    };
\addplot[width=0.15mm,color=violate,mark=square,mark size=0.5mm]
coordinates {
    (68.57, 99.75924)
    (81.365, 99.87176)
    (87.505, 99.91190)
    (90.745, 99.93273)
    (93.14, 99.94556)
    (94.60, 99.95421)
    (95.565, 99.96046)
    (96.12, 99.96523)
    (96.795, 99.96893)
    (97.23, 99.97191)
    (97.575, 99.97436)
    (97.875, 99.97641)
    (98.08, 99.97815)
    (98.245, 99.97965)
    (98.395, 99.98095)
    (98.51, 99.98209)
    (98.58, 99.98310)
    (98.64, 99.98400)
    (98.705, 99.98480)
    (98.745, 99.98553)
    };  
\end{axis}
\end{tikzpicture}
}
\subfloat[OpenAI-1536]{\vgap
\begin{tikzpicture}[scale=1]
\begin{axis}[
    height=\columnwidth/2.70,
    width=\columnwidth/2.0,
    xlabel=recall@20 (\%),
    ylabel=Prune-Rate (\%),
    label style={font=\scriptsize},
    tick label style={font=\scriptsize},
    xmin=90, xmax=99.7,
]
\addplot[width=0.15mm,color=amaranth,mark=o,mark size=0.5mm]
coordinates {
(86.295, 98.27271)
(90.77, 99.105009)
(93.03, 99.394444)
(94.46, 99.541261)
(95.31, 99.630774)
(96.085, 99.690588)
(96.57, 99.733887)
(96.98, 99.766646)
(97.335, 99.79198)
(97.555, 99.812399)
(97.72, 99.829127)
(97.925, 99.843171)
(98.12, 99.855014)
(98.31, 99.865169)
(98.435, 99.874003)
(98.57, 99.881761)
(98.675, 99.888609)
(98.78, 99.894717)
(98.86, 99.900182)
(98.895, 99.9051094)
};
\addplot[width=0.15mm,color=amber,mark=triangle,mark size=0.5mm]
coordinates {
    (86.45, 95.54127)
    (91.065, 97.57203)
    (93.34, 98.31827)
    (94.65, 98.70959)
    (95.585, 98.95117)
    (96.28, 99.11602)
    (96.76, 99.23534)
    (97.10, 99.32586)
    (97.42, 99.39732)
    (97.695, 99.45499)
    (97.845, 99.50250)
    (98.075, 99.54237)
    (98.245, 99.57619)
    (98.39, 99.60530)
    (98.595, 99.63069)
    (98.695, 99.65310)
    (98.795, 99.67283)
    (98.85, 99.69051)
    (98.93, 99.70625)
    (98.98, 99.72048)
    };
\addplot[width=0.15mm,color=violate,mark=square,mark size=0.5mm]
coordinates {
    (86.19, 99.7928)
    (90.805, 99.89612)
    (93.075, 99.92978)
    (94.38, 99.94691)
    (95.315, 99.95731)
    (96.00, 99.96429)
    (96.475, 99.96929)
    (96.815, 99.97307)
    (97.135, 99.97602)
    (97.41, 99.97839)
    (97.56, 99.98033)
    (97.79, 99.98195)
    (97.96, 99.98332)
    (98.10, 99.98450)
    (98.30, 99.98552)
    (98.395, 99.98642)
    (98.495, 99.98721)
    (98.55, 99.98792)
    (98.63, 99.98854)
    (98.68, 99.98911)
    };  
\end{axis}
\end{tikzpicture}
}
\vgap\caption{The In-Depth Analysis of Effectiveness}\label{fig:in-depth-analysis}
\end{small}\vspace{-1em}
\end{figure}

%% file: figures/parameter-select.tex
\begin{figure}[!t]
\centering
\begin{small}
\begin{tikzpicture}
    \begin{customlegend}[legend columns=3,
        legend entries={$\MRQ^{+}[64]$,$\MRQ^{+}[128]$,$\MRQ^{+}[256]$,$\MRQ^{+}[320]$,$\MRQ^{+}[384]$,$\MRQ^{+}[512]$},
        legend style={at={(0.45,1.15)},anchor=north,draw=none,font=\scriptsize,column sep=0.1cm}]
    \addlegendimage{line width=0.15mm,color=amber,mark=x,mark size=0.5mm}
    \addlegendimage{line width=0.2mm,color=violate,mark=o,mark size=0.5mm}
    \addlegendimage{line width=0.15mm,color=forestgreen,mark=triangle,mark size=0.5mm}
    \addlegendimage{line width=0.15mm,color=aliceblue,mark=square,mark size=0.5mm}
    \addlegendimage{line width=0.15mm,color=navy,mark=diamond,mark size=0.5mm}
    \addlegendimage{line width=0.2mm,color=amaranth,mark=pentagon,mark size=0.5mm}
    \end{customlegend}
\end{tikzpicture}
\\[-\lineskip]

\subfloat[GIST]{
\begin{tikzpicture}[scale=1]
\begin{axis}[
height=\columnwidth/2.70,
width=\columnwidth/1.90,
xlabel=recall@20,
ylabel=Qpsx100,
xmin=0.9,
xmax=1.005,
label style={font=\scriptsize},
title style={font=\scriptsize},
tick label style={font=\scriptsize},
ymajorgrids=true,
xmajorgrids=true,
grid style=dashed,
]
\addplot[line width=0.15mm,color=amber,mark=x,mark size=0.5mm,opacity=0.5]
plot coordinates {
    (0.7926, 25.5704)
    (0.8538, 22.4164)
    (0.8868, 19.5823)
    (0.9102, 17.8299)
    (0.9254, 16.1335)
    (0.9352, 15.5911)
    (0.9431, 13.6437)
    (0.9487, 13.6807)
    (0.9530, 13.0769)
    (0.95685, 12.4218)
    (0.95905, 11.7270)
    (0.96155, 11.3052)
    (0.9632, 11.1330)
    (0.9643, 10.9346)
    (0.96565, 10.4142)
    (0.9668, 10.2907)
    (0.9673, 9.94249)
    (0.96775, 9.75521)
    (0.96835, 9.33184)
};
\addplot[line width=0.2mm,color=violate,mark=o,mark size=0.5mm,opacity=1]
plot coordinates {
    (0.81365, 22.067)
    (0.87505, 19.2837)
    (0.90745, 17.5521)
    (0.9314, 14.8527)
    (0.95565, 13.3984)
    (0.9723, 13.0958)
    (0.97575, 12.1808)
    (0.97875, 11.4051)
    (0.9808, 11.0233)
    (0.98245, 11.2775)
    (0.98395, 10.9723)
    (0.9851, 10.3485)
    (0.9858, 10.0332)
    (0.9864, 9.74187)
    (0.98705, 9.19669)
    (0.98745, 8.93901)
};
\addplot[line width=0.15mm,color=forestgreen,mark=triangle,mark size=0.5mm,opacity=0.5]
plot coordinates {
    (0.82465, 16.2678)
    (0.97095, 12.2601)
    (0.97605, 12.3413)
    (0.9806, 11.3159)
    (0.9837, 11.0626)
    (0.9864, 10.7373)
    (0.9883, 10.238)
    (0.99005, 9.90765)
    (0.9913, 9.40155)
    (0.9924, 9.07849)
    (0.993, 8.6984)
    (0.9934, 8.29266)
    (0.99395, 8.24745)
    (0.9945, 7.97334)
};
\addplot[line width=0.15mm,color=aliceblue,mark=square,mark size=0.5mm,opacity=0.5]
plot coordinates {
    (0.82425, 19.3163)
    (0.8844, 17.1424)
    (0.9174, 14.769)
    (0.93715, 14.1382)
    (0.9523, 12.812)
    (0.963, 11.2992)
    (0.9708, 11.1398)
    (0.9762, 10.7009)
    (0.9808, 9.83533)
    (0.9838, 9.37244)
    (0.98685, 9.26822)
    (0.9885, 8.91463)
    (0.99005, 8.62408)
    (0.9913, 8.12826)
    (0.99255, 8.06547)
    (0.99345, 7.8437)
    (0.9942, 7.39267)
    (0.99475, 7.15259)
    (0.99505, 6.93837)
};
\addplot[line width=0.15mm,color=navy,mark=diamond,mark size=0.5mm,opacity=0.5]
plot coordinates {
    (0.8248, 12.3439)
    (0.88475, 14.3165)
    (0.91855, 12.5728)
    (0.9373, 13.0762)
    (0.95375, 11.9881)
    (0.9635, 11.4278)
    (0.9707, 10.6876)
    (0.9765, 9.82839)
    (0.98105, 9.23462)
    (0.9843, 8.83701)
    (0.9864, 8.32778)
    (0.98905, 8.11517)
    (0.99145, 7.61833)
    (0.99235, 7.43732)
    (0.99335, 7.1495)
    (0.99405, 6.90089)
    (0.99475, 6.23432)
    (0.9952, 6.19787)
    (0.9956, 6.10022)
};
\addplot[line width=0.15mm,color=amaranth,mark=pentagon,mark size=0.5mm,opacity=0.5]
plot coordinates {
    (0.69515, 5.11529)
    (0.82315, 13.2179)
    (0.884, 11.9092)
    (0.91805, 11.8187)
    (0.9379, 10.6759)
    (0.95345, 9.95205)
    (0.96365, 8.92771)
    (0.9718, 8.65778)
    (0.9775, 8.26796)
    (0.98165, 7.7355)
    (0.98485, 7.2651)
    (0.98715, 6.87928)
    (0.9891, 6.70063)
    (0.99075, 6.35442)
    (0.99165, 6.16953)
    (0.9927, 5.85335)
    (0.99325, 5.67479)
    (0.99425, 5.4129)
    (0.9946, 5.13984)
    (0.99525, 5.04096)
};
\end{axis}
\end{tikzpicture}
}
\subfloat[OpenAI-1536]{
\begin{tikzpicture}[scale=1]
\begin{axis}[
height=\columnwidth/2.70,
width=\columnwidth/1.90,
xlabel=recall@20,
ylabel=Qpsx100,
xmin=0.9,
xmax=0.99,
label style={font=\scriptsize},
title style={font=\scriptsize},
tick label style={font=\scriptsize},
ymajorgrids=true,
xmajorgrids=true,
grid style=dashed,
]
\addplot[line width=0.15mm,color=violate,mark=o,mark size=0.5mm,opacity=0.6]
plot coordinates {
    (0.8432, 2.92796)
    (0.8969, 2.59067)
    (0.922, 2.22298)
    (0.93845, 2.02132)
    (0.94915, 1.79905)
    (0.95605, 1.64602)
    (0.9622, 1.52181)
    (0.96755, 1.40521)
    (0.9716, 1.29117)
    (0.9739, 1.31564)
    (0.9769, 1.28635)
    (0.9786, 1.22372)
    (0.98045, 1.18515)
    (0.9817, 1.14119)
    (0.98285, 1.11487)
    (0.9835, 1.08746)
    (0.98465, 1.06032)
    (0.98555, 1.02266)
    (0.9863, 1.00687)
    (0.987, 0.989576)
};
\addplot[line width=0.15mm,color=forestgreen,mark=triangle,mark size=0.5mm,opacity=0.6]
plot coordinates {
    (0.9064, 8.89399)
    (0.93045, 7.85498)
    (0.9453, 7.111)
    (0.9531, 6.74823)
    (0.96025, 6.22847)
    (0.96535, 5.91294)
    (0.9692, 5.92708)
    (0.97185, 5.61972)
    (0.97465, 5.3954)
    (0.97755, 5.16205)
    (0.9795, 5.03633)
    (0.9814, 4.87958)
    (0.9828, 4.77395)
    (0.98435, 4.63617)
    (0.98515, 4.51872)
    (0.9864, 4.38863)
    (0.98735, 4.31483)
    (0.98815, 4.18056)
    (0.9887, 4.08739)
};
\addplot[line width=0.15mm,color=aliceblue,mark=square,mark size=0.5mm,opacity=0.6]
plot coordinates {
    (0.9035, 14.8709)
    (0.92565, 13.5697)
    (0.93935, 12.5301)
    (0.9631, 9.26225)
    (0.9658, 8.99667)
    (0.968, 8.0732)
    (0.974, 7.35586)
    (0.9755, 7.21068)
    (0.97715, 6.79286)
    (0.9786, 6.55632)
    (0.9794, 6.2974)
    (0.98005, 6.03154)
    (0.98045, 5.97088)
    (0.98095, 5.93725)
};
\addplot[line width=0.15mm,color=navy,mark=diamond,mark size=0.5mm,opacity=0.6]
plot coordinates {
    (0.906, 14.7249)
    (0.928, 13.1016)
    (0.94135, 12.2701)
    (0.94985, 11.1529)
    (0.95625, 10.3907)
    (0.9619, 9.71178)
    (0.96535, 9.1593)
    (0.96845, 8.90802)
    (0.97065, 8.48693)
    (0.97245, 8.0105)
    (0.97435, 7.66101)
    (0.9757, 7.33115)
    (0.9775, 6.99072)
    (0.97865, 6.7015)
    (0.98005, 6.53156)
    (0.9811, 6.3691)
    (0.98255, 6.12057)
    (0.98315, 5.93097)
    (0.9834, 5.73885)
};
\addplot[line width=0.15mm,color=amaranth,mark=pentagon,mark size=0.5mm,opacity=1]
plot coordinates {
    (0.9085, 15.0822)
    (0.93065, 13.4055)
    (0.9442, 11.824)
    (0.95325, 10.7564)
    (0.9603, 9.75942)
    (0.96495, 9.41934)
    (0.9685, 8.55656)
    (0.9716, 7.62221)
    (0.9744, 7.16239)
    (0.97585, 7.23156)
    (0.978, 7.10973)
    (0.9797, 6.63727)
    (0.9812, 6.39644)
    (0.9829, 5.77026)
    (0.984, 5.79095)
    (0.985, 5.62728)
    (0.9857, 5.29309)
    (0.98665, 5.17639)
    (0.987, 4.93211)
};
\end{axis}
\end{tikzpicture}
}

\vgap\caption{The Validation of Parameters Selection}\label{fig:parameter-select}\vgap
\end{small}
\end{figure}

%% file: appendix.tex
\section*{A. Appendix}
\subsection*{A.1. Extra Related Work}

\subsection{Re-Ranking}\label{sub:re-rank}
One limitation of quantization techniques is the reduction in search accuracy (i.e., recall) due to approximate distance computation.
To mitigate this issue, a common strategy is to apply re-ranking using exact distances~\cite{Diskann-NIPS-2019,ADSampling:journals/sigmod/GaoL23,Rabitq-SIGMOD-2024,ExRaBitQ-arxiv-2024,Starling-SIGMOD-2024-Mengzhao}.
Specifically, the method first performs AKNN search using approximate distances and retains a larger candidate set, such as the top-$R$ results (where $R > k$).
Then, it re-computes the exact distances for all candidates and selects the final top-$K$ nearest neighbors based on these corrected values.
The re-rank method is widely used in disk-based AKNN index~\cite{Diskann-NIPS-2019,Starling-SIGMOD-2024-Mengzhao}, where compressed vectors are stored in memory and full-precision vectors are stored on disk. When obtaining a temporary Top $R$ nearest neighbor with approximate distance, the re-rank process loads the exact vectors from the disk and recomputes the distance to improve search accuracy.

\stitle{Error Bounds.}
Re-ranking introduces additional overhead because it requires many exact distance computations. 
To alleviate this cost, recent AKNN methods such as RabitQ~\cite{Rabitq-SIGMOD-2024} incorporate error bounds to prune unlikely candidates early. 
By exploiting statistical properties of the quantization process, RabitQ can determine whether the true distance of a candidate is guaranteed to exceed that of the current $k$-th nearest neighbor. 
If so, the candidate can be safely discarded without performing an exact distance computation. 
However, as discussed in the Introduction, this optimization is inherently tied to the constraint that the code length must equal the vector dimensionality, limiting its applicability. 
Our work, $\MRQ$, aims to retain the benefits of error-bounded re-ranking while removing this restrictive constraint.

\sstitle{Discussion.}
Methods for improving AKNN search accuracy through distance correction, re-ranking, and error-based approaches essentially aim to obtain the precise distances to the k-nearest neighbors of the query. The key difference is that the method based on error bound provides guarantees on search accuracy, whereas the re-ranking method requires empirically setting the re-ranking range to achieve the desired AKNN search accuracy target.

\subsection*{A.2. Extra Discussion}
\begin{figure*}[t]
\centering
\begin{small}
\subfloat[GIST]{
\includegraphics[width=0.495\columnwidth]{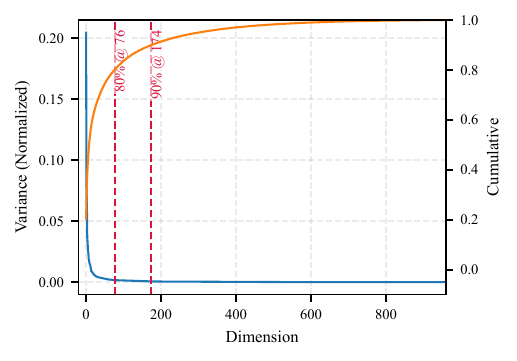}
}
\subfloat[MSONG]{
\includegraphics[width=0.495\columnwidth]{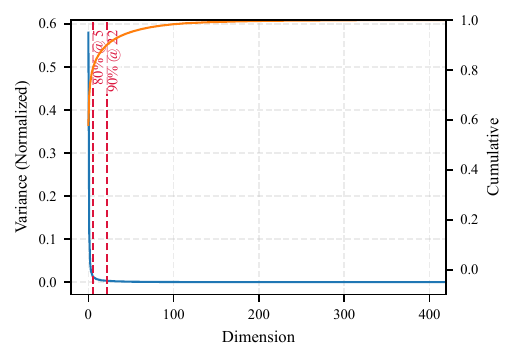}
}
\subfloat[OpenAI-1536]{
\includegraphics[width=0.495\columnwidth]{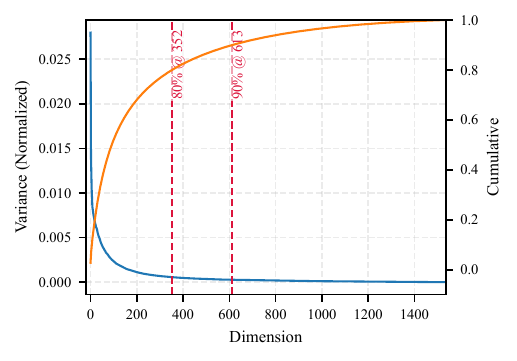}
}
\subfloat[OpenAI-3072]{
\includegraphics[width=0.495\columnwidth]{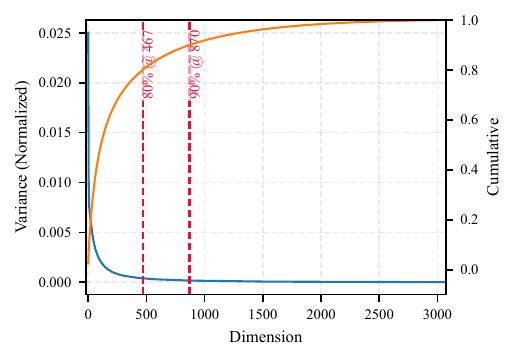}
}
\\
\subfloat[DEEP]{
\includegraphics[width=0.495\columnwidth]{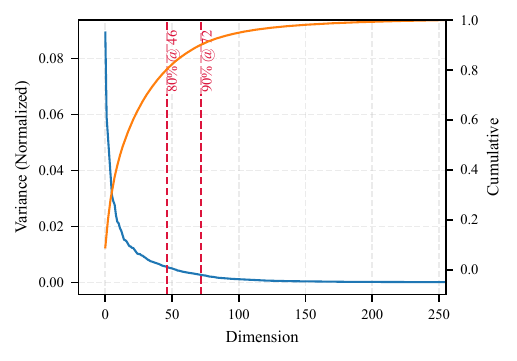}
}
\subfloat[WORD2VEC]{
\includegraphics[width=0.495\columnwidth]{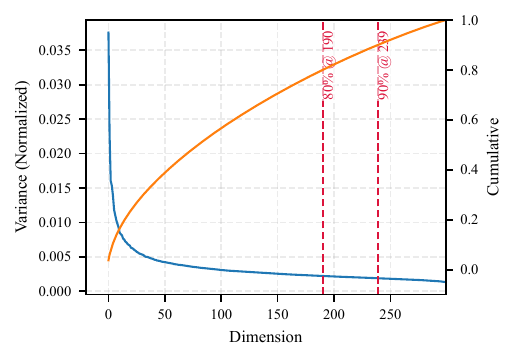}
}
\subfloat[MSMARCO]{
\includegraphics[width=0.495\columnwidth]{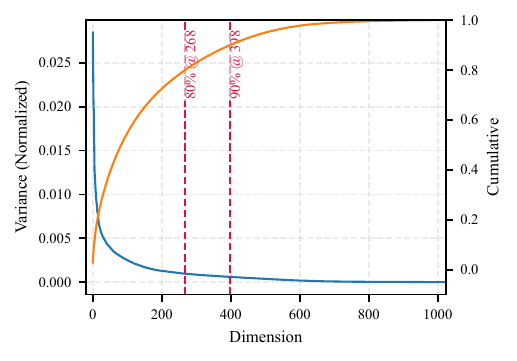}
}
\subfloat[TINY]{
\includegraphics[width=0.495\columnwidth]{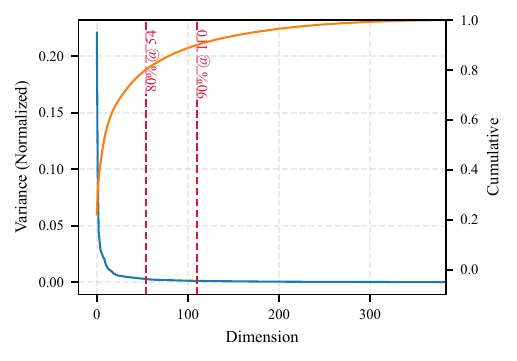}
}
\caption{The Variance Distributions of Vectors After PCA}\label{fig:vec_var_all}
\end{small}
\end{figure*}

We further evaluate the variance statistics of several representative datasets. 
As shown in Fig.~\ref{fig:vec_var_all}, vector embeddings from images, text, and audio consistently exhibit a long-tailed variance distribution after rotation by the PCA matrix. 
However, the cumulative variance captured by the tail dimensions differs across data modalities. 
For example, \textsc{Word2Vec} captures less than $80\%$ of the total variance using the top $50\%$ of its dimensions, whereas \textsc{GIST} reaches over $80\%$ using only $10\%$ of its dimensions.

These observations align with our empirical parameter-setting rule in Eq.~\ref{eq:para-set}. 
The projection dimensions required to preserve $80\%$ of the total variance for the \textsc{GIST}, \textsc{MSONG}, \textsc{OpenAI-1536}, \textsc{OpenAI-3072}, \textsc{DEEP}, \textsc{Word2Vec}, \textsc{MSMARCO}, and \textsc{TINY} datasets are $128$, $5$, $354$, $470$, $46$, $173$, $273$, and $51$, respectively. 
Following this rule, we set the projected dimensionality to $128$ bits for \textsc{MSONG}, \textsc{DEEP}, \textsc{TINY}, and \textsc{GIST}; $256$ bits for \textsc{Word2Vec}; and $512$ bits for \textsc{OpenAI-1536}, \textsc{OpenAI-3072}, and \textsc{MSMARCO10M}, achieving near-optimal search efficiency across all datasets.

\input{figures/ablation}
\input{figures/MR-PQ-SQ}

\input{figures/extra-quantiztaion}

\subsection*{A.2. Extra Experiments}

\stitle{Exp-5: Boost PQ and SQ query Efficiency.}
The projection-and-vector-split strategy can also enhance the query efficiency of mainstream codebook-based and dimension-wise quantizers beyond RabitQ. We evaluate the generalizability of $\MRQ$'s core concept using OPQ and 8-bit SQ—methods commonly used in practice—denoted as MR-PQ and MR-SQ, respectively. We use empirical formulas to set the projection dimensions, keeping other quantization parameters such as the number of bits per dimension (subspace) unchanged. Since OPQ and SQ lack a probabilistic error bound, we fix a reranking range (e.g., 500) to prevent degradation in search accuracy.

As shown in Fig.~\ref{fig:MR-PQ-SQ}, the MR-style variants consistently shift the performance curve upward, achieving higher QPS at comparable recall levels relative to their original counterparts. This improvement stems from the concentration of variance in the leading dimensions following PCA; quantizing only this informative subspace reduces both (i) the computation required per candidate and (ii) memory traffic during scanning. Meanwhile, the residual error has a negligible impact on overall distance estimation, thereby avoiding significant drops in search accuracy.

\stitle{Exp-6: Study of Centroid Approximation.}
In Algorithm~\ref{algo:MRQ-Index}, we apply the k-means algorithm on the projected (approximate) vectors instead of the original full-dimensional vectors. This results in approximate centroids, as used in the IVF method. The effectiveness of this approach is shown in Fig.~\ref{fig:index-app}, where the x-axis indicates the number of scanned clusters. The red line labeled $\IVF$ represents the original centroids computed from the full-dimensional vectors, while the blue line labeled $\IVF^{*}$ represents the approximate centroids computed from the projected vectors.
Specifically, $\IVF$ uses all $D$ dimensions to compute the centroids, whereas $\IVF^{*}$ uses only 128 dimensions for the GIST dataset (1/7 of the original dimensionality) and 512 dimensions for the OpenAI-1536 dataset (1/3 of the original dimensionality).

As shown in Fig.~\ref{fig:index-app}, the projected centroids incur almost no recall loss on the GIST dataset and even slightly outperform the original centroids on the OpenAI-1536 dataset. Therefore, we adopt the projected centroids in our method. Additionally, this projection significantly reduces the training time for the IVF index, making it more efficient for large-scale datasets.

\stitle{Exp-7: Comparison with Additional Baselines.}
To further validate the superiority of our proposed quantization method $\MRQ$, we first disable the projection-distance and memory-layout optimizations. 
(1) We also include additional quantization techniques, such as Product Quantization (PQ), as baselines. 
For a fair comparison in terms of memory usage, IVF-PQ is configured with a $32\times$ compression ratio so that its space consumption matches that of IVF-RabitQ. 
Note that both IVF-PQ and IVF-\textsc{Truncate} are evaluated without re-ranking.
(2) Under the re-ranking setting, we additionally compare against the state-of-the-art Binary Quantization (BQ) method RabitQ, as well as the optimized PQ variant, OPQ-4bit-fastscan (denoted IVF-OPQfsr) with the best parameter set in~\cite{Rabitq-SIGMOD-2024}. 
(3) We also design an IVF-\textsc{Truncate} baseline with re-ranking, which directly truncates vectors to match the bit budget of $\MRQ$. 
This method is similar in spirit to $\MRQ$ but lacks PCA, and thus cannot flexibly control the separation of leading and residual dimensions. 
All results are reported in Fig.~\ref{fig:extra-quantiztaion}.

As shown in Fig.~\ref{fig:extra-quantiztaion}, without re-ranking, standard PQ-based quantization fails to achieve $90\%$ recall across datasets. 
Using fewer bits—comparable to the setting of $\MRQ$—only further degrades accuracy.
With re-ranking enabled, the results show that $\MRQ$ significantly outperforms both IVF-RabitQ and IVF-OPQfsr on the vast majority of datasets. 
Moreover, compared with \textsc{Truncate}, $\MRQ$ exhibits a substantial performance advantage, demonstrating the importance of PCA-based projection. 
Notably, $\MRQ$ achieves this superior accuracy relying solely on quantized distances for re-ranking, even without multi-stage distance computation or memory-layout optimizations.




%% file: figures/ablation.tex
\begin{figure}[h]
\centering
\begin{small}
\begin{tikzpicture}
    \begin{customlegend}[legend columns=2,
        legend entries={$\IVF$,$\IVF^{*}$},
        legend style={at={(0.45,1.15)},anchor=north,draw=none,font=\scriptsize,column sep=0.1cm}]
    \addlegendimage{line width=0.15mm,color=amaranth,mark=o,mark size=0.5mm}
    \addlegendimage{line width=0.15mm,color=navy,mark=triangle,mark size=0.5mm}
    \end{customlegend}
\end{tikzpicture}
\\[-\lineskip]

\subfloat[GIST]{

\begin{tikzpicture}[scale=1]
\begin{axis}[
    height=\columnwidth/2.60,
width=\columnwidth/1.90,
ylabel=recall@20,
xlabel=nprobe,
label style={font=\scriptsize},
tick label style={font=\scriptsize},
ymajorgrids=true,
xmajorgrids=true,
grid style=dashed,
]
\addplot[line width=0.15mm,color=amaranth,mark=o,mark size=0.5mm]
plot coordinates {
    ( 25.0, 0.701 )
    ( 50.0, 0.828 )
    ( 75.0, 0.888 )
    ( 100.0, 0.92 )
    ( 125.0, 0.942 )
    ( 150.0, 0.955 )
    ( 175.0, 0.966 )
    ( 200.0, 0.973 )
    ( 225.0, 0.979 )
    ( 250.0, 0.983 )
    ( 275.0, 0.987 )
    ( 300.0, 0.989 )
    ( 325.0, 0.992 )
    ( 350.0, 0.993 )
    ( 375.0, 0.994 )
    ( 400.0, 0.995 )
    ( 425.0, 0.995 )
    ( 450.0, 0.996 )
    ( 475.0, 0.997 )
    ( 500.0, 0.997 )
};
\addplot[line width=0.15mm,color=navy,mark=triangle,mark size=0.5mm]
plot coordinates {
    ( 25.0, 0.689 )
    ( 50.0, 0.819 )
    ( 75.0, 0.882 )
    ( 100.0, 0.914 )
    ( 125.0, 0.937 )
    ( 150.0, 0.953 )
    ( 175.0, 0.963 )
    ( 200.0, 0.969 )
    ( 225.0, 0.975 )
    ( 250.0, 0.98 )
    ( 275.0, 0.983 )
    ( 300.0, 0.987 )
    ( 325.0, 0.989 )
    ( 350.0, 0.991 )
    ( 375.0, 0.993 )
    ( 400.0, 0.994 )
    ( 425.0, 0.995 )
    ( 450.0, 0.995 )
    ( 475.0, 0.996 )
    ( 500.0, 0.997 )
};

\end{axis}
\end{tikzpicture}\hspace{2mm}
}
\subfloat[OpenAI-1536]{

\begin{tikzpicture}[scale=1]
\begin{axis}[
    height=\columnwidth/2.60,
width=\columnwidth/1.90,
ylabel=recall@20,
xlabel=nprobe,
label style={font=\scriptsize},
tick label style={font=\scriptsize},
ymajorgrids=true,
xmajorgrids=true,
grid style=dashed,
]
\addplot[line width=0.15mm,color=amaranth,mark=o,mark size=0.5mm]
plot coordinates {
    ( 30.0, 0.863 )
    ( 60.0, 0.908 )
    ( 90.0, 0.931 )
    ( 120.0, 0.945 )
    ( 150.0, 0.953 )
    ( 180.0, 0.961 )
    ( 210.0, 0.966 )
    ( 240.0, 0.97 )
    ( 270.0, 0.974 )
    ( 300.0, 0.976 )
    ( 330.0, 0.978 )
    ( 360.0, 0.98 )
    ( 390.0, 0.982 )
    ( 420.0, 0.984 )
    ( 450.0, 0.985 )
    ( 480.0, 0.986 )
    ( 510.0, 0.987 )
    ( 540.0, 0.988 )
    ( 570.0, 0.989 )
    ( 600.0, 0.989 )
};
\addplot[line width=0.15mm,color=navy,mark=triangle,mark size=0.5mm]
plot coordinates {
    ( 30.0, 0.866 )
    ( 60.0, 0.912 )
    ( 90.0, 0.935 )
    ( 120.0, 0.948 )
    ( 150.0, 0.957 )
    ( 180.0, 0.964 )
    ( 210.0, 0.969 )
    ( 240.0, 0.972 )
    ( 270.0, 0.975 )
    ( 300.0, 0.978 )
    ( 330.0, 0.98 )
    ( 360.0, 0.982 )
    ( 390.0, 0.984 )
    ( 420.0, 0.985 )
    ( 450.0, 0.987 )
    ( 480.0, 0.988 )
    ( 510.0, 0.989 )
    ( 540.0, 0.99 )
    ( 570.0, 0.991 )
    ( 600.0, 0.991 )
};

\end{axis}
\end{tikzpicture}\hspace{2mm}
}

\vgap\caption{The Study of Centroid Approximation}\vgap\vgap\label{fig:index-app}
\end{small}
\end{figure}

%% file: figures/MR-PQ-SQ.tex
\begin{figure}[htbp]
\centering
\begin{small}
\begin{tikzpicture}
    \begin{customlegend}[legend columns=4,
        legend entries={OPQ, MR-OPQ, SQ, MR-SQ},
        legend style={at={(0.45,1.15)},anchor=north,draw=none,font=\scriptsize,column sep=0.1cm}]
    \addlegendimage{line width=0.15mm,color=amaranth,mark=o,mark size=0.5mm}
    \addlegendimage{line width=0.15mm,color=violate,mark=triangle,mark size=0.5mm}
    \addlegendimage{line width=0.15mm,color=amber,mark=square,mark size=0.5mm}
    \addlegendimage{line width=0.15mm,color=navy,mark=diamond,mark size=0.5mm}
    \end{customlegend}
\end{tikzpicture}
\\[-\lineskip]

\subfloat[GIST-PQ]{\vgap
\begin{tikzpicture}[scale=1]
\begin{axis}[
height=\columnwidth/2.70,
width=\columnwidth/1.90,
xlabel=recall@20,
ylabel=Qpsx100,
xmax=1.005,
label style={font=\scriptsize},
title style={font=\scriptsize},
tick label style={font=\scriptsize},
ymajorgrids=true,
xmajorgrids=true,
grid style=dashed,
]
\addplot[line width=0.15mm,color=amaranth,mark=o,mark size=0.5mm]
plot coordinates {
(0.65775, 3.26)
(0.78975, 3.56075)
(0.85575, 3.16967)
(0.8967, 2.72806)
(0.9208, 2.37825)
(0.93865, 2.12229)
(0.9516, 1.91047)
(0.95965, 1.74205)
(0.9671, 1.59935)
(0.973, 1.45888)
(0.97795, 1.31446)
(0.98195, 1.23638)
(0.9848, 1.18003)
(0.9872, 1.12195)
(0.9894, 1.04466)
(0.9909, 1.0111)
(0.99265, 0.970172)
(0.9933, 0.925623)
(0.99435, 0.889711)
(0.99475, 0.848912)
};
\addplot[line width=0.15mm,color=violate,mark=triangle,mark size=0.5mm]
plot coordinates {
(0.64325, 12.5939)
(0.7802, 10.9279)
(0.84755, 9.90078)
(0.8868, 9.24374)
(0.91045, 8.66515)
(0.93065, 8.13083)
(0.94255, 7.72737)
(0.9523, 7.32234)
(0.95875, 6.96415)
(0.96255, 6.68398)
(0.9674, 6.42893)
(0.97155, 6.13909)
(0.97375, 5.93293)
(0.9762, 5.73636)
(0.97835, 5.51882)
(0.98005, 5.33313)
(0.98135, 5.1798)
(0.98215, 5.03494)
(0.98345, 4.87577)
(0.9839, 4.72258)
};
\end{axis}
\end{tikzpicture}
}
\subfloat[GIST-SQ]{\vgap
\begin{tikzpicture}[scale=1]
\begin{axis}[
height=\columnwidth/2.70,
width=\columnwidth/1.90,
xlabel=recall@20,
ylabel=Qpsx100,
label style={font=\scriptsize},
title style={font=\scriptsize},
tick label style={font=\scriptsize},
ymajorgrids=true,
xmajorgrids=true,
grid style=dashed,
]
\addplot[line width=0.15mm,color=amber,mark=square,mark size=0.5mm]
plot coordinates {
(0.65775, 0.636489)
(0.78975, 0.327931)
(0.85575, 0.220927)
(0.8967, 0.166791)
(0.9208, 0.134615)
(0.93865, 0.112623)
(0.9516, 0.0968985)
(0.95965, 0.0851073)
(0.9671, 0.0760718)
(0.973, 0.0685383)
(0.97795, 0.0625637)
(0.98195, 0.0574751)
(0.9848, 0.0533409)
(0.9872, 0.0496354)
(0.9894, 0.0465155)
(0.9909, 0.0436309)
(0.99265, 0.0412315)
(0.9933, 0.0389767)
(0.99435, 0.0369726)
(0.99475, 0.0352899)
};
\addplot[line width=0.15mm,color=navy,mark=diamond,mark size=0.5mm]
plot coordinates {
(0.6435, 4.62796)
(0.7808, 2.47503)
(0.84935, 1.68981)
(0.88955, 1.2999)
(0.9142, 1.05362)
(0.9352, 0.887907)
(0.94765, 0.761789)
(0.95765, 0.673254)
(0.96445, 0.602707)
(0.96865, 0.543379)
(0.97425, 0.493723)
(0.97865, 0.453921)
(0.98125, 0.423107)
(0.9839, 0.393404)
(0.98635, 0.369184)
(0.9882, 0.346637)
(0.9896, 0.327123)
(0.9907, 0.309699)
(0.99225, 0.294045)
(0.993, 0.280234)
};
\end{axis}
\end{tikzpicture}
}
\\
\subfloat[OpenAI-PQ]{\vgap
\begin{tikzpicture}[scale=1]
\begin{axis}[
height=\columnwidth/2.70,
width=\columnwidth/1.90,
xlabel=recall@20,
ylabel=Qpsx100,
xmax=1.005,
label style={font=\scriptsize},
title style={font=\scriptsize},
tick label style={font=\scriptsize},
ymajorgrids=true,
xmajorgrids=true,
grid style=dashed,
]
\addplot[line width=0.15mm,color=amaranth,mark=o,mark size=0.5mm]
plot coordinates {
(0.8309, 3.0501)
(0.8809, 2.53334)
(0.90825, 2.16825)
(0.9242, 1.90018)
(0.9366, 1.69547)
(0.94515, 1.5253)
(0.951, 1.37177)
(0.9562, 1.27589)
(0.9613, 1.17792)
(0.96455, 1.0943)
(0.96745, 1.02221)
(0.9702, 0.960466)
(0.9729, 0.903972)
(0.9745, 0.854026)
(0.9761, 0.808927)
(0.9771, 0.769156)
(0.97845, 0.732502)
(0.9799, 0.699129)
(0.9812, 0.669476)
(0.98225, 0.640643)
};
\addplot[line width=0.15mm,color=violate,mark=triangle,mark size=0.5mm]
plot coordinates {
(0.8359, 6.70534)
(0.8852, 6.00926)
(0.91185, 5.3872)
(0.9289, 4.86778)
(0.93975, 4.4823)
(0.94775, 4.15693)
(0.9543, 3.87661)
(0.96025, 3.62466)
(0.964, 3.40456)
(0.9673, 3.20964)
(0.96985, 3.03967)
(0.9722, 2.88258)
(0.9743, 2.74959)
(0.97665, 2.6203)
(0.9782, 2.50271)
(0.9794, 2.39676)
(0.98035, 2.30305)
(0.982, 2.21248)
(0.9832, 2.13053)
(0.98425, 2.05488)
};
\end{axis}
\end{tikzpicture}
}
\subfloat[OpenAI-SQ]{\vgap
\begin{tikzpicture}[scale=1]
\begin{axis}[
height=\columnwidth/2.70,
width=\columnwidth/1.90,
xlabel=recall@20,
ylabel=Qpsx100,
label style={font=\scriptsize},
title style={font=\scriptsize},
tick label style={font=\scriptsize},
ymajorgrids=true,
xmajorgrids=true,
grid style=dashed,
]
\addplot[line width=0.15mm,color=amber,mark=square,mark size=0.5mm]
plot coordinates {
(0.8309, 0.502925)
(0.8809, 0.267626)
(0.90825, 0.185154)
(0.9242, 0.140272)
(0.9366, 0.113168)
(0.94515, 0.0947357)
(0.951, 0.0816027)
(0.9562, 0.0716288)
(0.9613, 0.0638162)
(0.96455, 0.057543)
(0.96745, 0.0524174)
(0.9702, 0.0480845)
(0.9729, 0.0444801)
(0.9745, 0.0413215)
(0.9761, 0.0386023)
(0.9771, 0.0362128)
(0.97845, 0.0340051)
(0.9799, 0.0322138)
(0.9812, 0.0264739)
(0.98225, 0.0290407)
};
\addplot[line width=0.15mm,color=navy,mark=diamond,mark size=0.5mm]
plot coordinates {
(0.8359, 1.50749)
(0.8852, 0.766266)
(0.91185, 0.569804)
(0.9289, 0.431857)
(0.93975, 0.348026)
(0.94775, 0.291979)
(0.9543, 0.249693)
(0.9603, 0.216402)
(0.96405, 0.194661)
(0.96735, 0.176497)
(0.9699, 0.160496)
(0.97225, 0.147533)
(0.97435, 0.136588)
(0.9767, 0.126911)
(0.97825, 0.118759)
(0.97945, 0.111377)
(0.9804, 0.104944)
(0.98205, 0.099125)
(0.98325, 0.0938548)
(0.9843, 0.089489)
};
\end{axis}
\end{tikzpicture}
}

\vgap\caption{Extend to Product and Scalar Quantization Methods}\label{fig:MR-PQ-SQ}\vgap
\end{small}
\end{figure}

%% file: figures/extra-quantiztaion.tex
\begin{figure*}[t]
\centering
\begin{small}
\begin{tikzpicture}
    \begin{customlegend}[legend columns=6,
        legend entries={$\IVF$-RabitQ,$\IVF$-$\MRQ$,$\IVF$-PQ,$\IVF$-OPQfsr,$\IVF$-\textsc{Truncate},$\IVF$-RabitQ$^{+}$},
        legend style={at={(0.45,1.15)},anchor=north,draw=none,font=\scriptsize,column sep=0.1cm}]
    \addlegendimage{line width=0.15mm,color=amaranth,mark=o,mark size=0.5mm}
    \addlegendimage{line width=0.15mm,color=amber,mark=triangle,mark size=0.5mm}
    \addlegendimage{line width=0.15mm,color=black,mark=square,mark size=0.5mm}
    \addlegendimage{line width=0.15mm,color=navy,mark=oplus,mark size=0.5mm}
    \addlegendimage{line width=0.15mm,color=black,mark=x,mark size=0.5mm}
    \addlegendimage{line width=0.15mm,color=violate,mark=diamond,mark size=0.5mm}
    \end{customlegend}
\end{tikzpicture}
\\[-\lineskip]

\subfloat[GIST]{\vspace{-2mm}
\begin{tikzpicture}[scale=1]
\begin{axis}[
    height=\columnwidth/2.70,
width=\columnwidth/1.90,
xlabel=recall@20,
ylabel=Qpsx100,
label style={font=\scriptsize},
tick label style={font=\scriptsize},
ymajorgrids=true,
xmajorgrids=true,
grid style=dashed,
]
\addplot[line width=0.15mm,color=amaranth,mark=o,mark size=0.5mm]
plot coordinates {
(0.70025, 11.0196)
(0.8272, 9.48403)
(0.8869, 8.38973)
(0.9189, 7.48001)
(0.94025, 6.75347)
(0.9537, 6.28338)
(0.9638, 5.8112)
(0.97095, 5.40643)
(0.9776, 4.99749)
(0.9816, 4.74842)
(0.98475, 4.1899)
(0.98725, 4.18877)
(0.9896, 3.95406)
(0.9912, 3.78924)
(0.99205, 3.62139)
(0.99295, 3.43964)
(0.9935, 3.30898)
(0.9943, 3.17762)
(0.99495, 3.07247)
(0.99535, 2.9571)
};
\addplot[line width=0.15mm,color=amber,mark=triangle,mark size=0.5mm]
plot coordinates {
    ( 0.689, 20.532 )
    ( 0.817, 16.099 )
    ( 0.881, 13.534 )
    ( 0.914, 12.24 )
    ( 0.937, 11.111 )
    ( 0.952, 10.318 )
    ( 0.962, 9.79 )
    ( 0.968, 9.285 )
    ( 0.974, 8.85 )
    ( 0.979, 8.447 )
    ( 0.982, 8.071 )
    ( 0.986, 7.874 )
    ( 0.987, 7.526 )
    ( 0.989, 7.358 )
    ( 0.991, 7.054 )
    ( 0.992, 6.884 )
    ( 0.994, 6.574 )
    ( 0.995, 6.344 )
};
\addplot[line width=0.15mm,color=navy,mark=oplus,mark size=0.5mm]
plot coordinates {
    ( 0.77475, 9.391 )
    ( 0.8859, 6.304 )
    ( 0.9575, 3.860 )
    ( 0.9895, 2.171 )
    ( 0.99825, 1.170 )
};
\addplot[line width=0.15mm,color=black,mark=square,mark size=0.5mm]
plot coordinates {
    ( 0.46745, 11.320332054189303 )
    ( 0.51855, 6.029535062499442 )
    ( 0.54595, 2.9525761839987916 )
    ( 0.5574, 1.5682529666858315 )
    ( 0.56055, 0.8137309122083259 )
    ( 0.5613, 0.4242786641099875 )
};
\addplot[line width=0.15mm,color=black,mark=x,mark size=0.5mm]
plot coordinates {
(0.94175, 4.03645)
(0.96785, 2.67131)
(0.97985, 2.04611)
(0.98555, 1.65488)
(0.98975, 1.3923)
(0.99235, 1.20094)
(0.9938, 1.06832)
(0.99535, 0.95318)
(0.99635, 0.869266)
(0.9967, 0.79894)
(0.99715, 0.742706)
(0.99735, 0.690741)
(0.99775, 0.649583)
(0.99795, 0.614293)
(0.99805, 0.582423)
(0.9983, 0.554952)
(0.9985, 0.530866)
(0.9986, 0.510026)
(0.9988, 0.492162)
};
\addplot[line width=0.15mm,color=violate,mark=diamond,mark size=0.5mm]
plot coordinates {
(0.8282, 9.53588)
(0.88755, 6.20761)
(0.9191, 6.97749)
(0.94045, 6.62243)
(0.95425, 6.09919)
(0.9646, 5.68378)
(0.97175, 5.25812)
(0.9783, 4.84151)
(0.98225, 4.59115)
(0.9855, 4.33362)
(0.9881, 4.09946)
(0.99025, 3.88943)
(0.99175, 3.70667)
(0.99285, 3.55998)
(0.99365, 3.37721)
(0.9942, 3.25256)
(0.99495, 3.0953)
(0.9958, 2.92044)
(0.9961, 2.8038)
};
\end{axis}
\end{tikzpicture}\hspace{2mm}
}
\subfloat[MSONG]{\vspace{-2mm}
\begin{tikzpicture}[scale=1]
\begin{axis}[
    height=\columnwidth/2.70,
width=\columnwidth/1.90,
xlabel=recall@20,
ylabel=Qpsx1000,
label style={font=\scriptsize},
tick label style={font=\scriptsize},
ymajorgrids=true,
xmajorgrids=true,
grid style=dashed,
]
\addplot[line width=0.15mm,color=amaranth,mark=o,mark size=0.5mm]
plot coordinates {
    ( 0.752, 2.991 )
    ( 0.886, 2.936 )
    ( 0.933, 2.799 )
    ( 0.957, 2.61 )
    ( 0.97, 2.571 )
    ( 0.978, 2.482 )
    ( 0.983, 2.398 )
    ( 0.987, 2.359 )
    ( 0.989, 2.201 )
    ( 0.991, 2.184 )
    ( 0.993, 2.137 )
    ( 0.994, 2.061 )
    ( 0.996, 1.967 )
    ( 0.997, 1.85 )
    ( 0.998, 1.77 )
};
\addplot[line width=0.15mm,color=amber,mark=triangle,mark size=0.5mm]
plot coordinates {
    ( 0.749, 6.862 )
    ( 0.884, 6.109 )
    ( 0.934, 5.734 )
    ( 0.956, 5.135 )
    ( 0.97, 4.868 )
    ( 0.977, 4.557 )
    ( 0.983, 4.427 )
    ( 0.986, 4.208 )
    ( 0.99, 4.208 )
    ( 0.992, 4.004 )
    ( 0.993, 3.87 )
    ( 0.994, 3.711 )
    ( 0.995, 3.548 )
    ( 0.997, 3.376 )
    ( 0.998, 3.064 )
};
\addplot[line width=0.15mm,color=black,mark=square,mark size=0.5mm]
plot coordinates {
    ( 0.545, 7.281448223441682 )
    ( 0.59935, 4.561754794349616 )
    ( 0.62625, 2.6244032900619695 )
    ( 0.6344, 1.4564295765783743 )
    ( 0.6369, 0.768050936646156 )
    ( 0.6375, 0.39953419059842605 )
    ( 0.63765, 0.20640399993737455 )
    ( 0.63765, 0.10811816197507339 )
};
\addplot[line width=0.15mm,color=black,mark=x,mark size=0.5mm]
plot coordinates {
(0.81, 2.01867)
(0.8684, 1.40513)
(0.89705, 0.993694)
(0.914, 0.79146)
(0.92585, 0.692995)
(0.934, 0.609347)
(0.9403, 0.540996)
(0.945, 0.499297)
(0.949, 0.449896)
(0.9522, 0.421748)
(0.9544, 0.37914)
(0.95615, 0.35155)
(0.9577, 0.330106)
(0.95885, 0.307489)
(0.96015, 0.290307)
(0.96085, 0.273115)
(0.96155, 0.258972)
(0.9622, 0.244751)
(0.9627, 0.234249)
};
\addplot[line width=0.15mm,color=violate,mark=diamond,mark size=0.5mm]
plot coordinates {
    (0.8836, 3.24849)
    (0.9313, 3.07324)
    (0.9543, 2.87205)
    (0.9676, 2.79641)
    (0.97565, 2.65278)
    (0.98115, 2.56295)
    (0.98455, 2.47136)
    (0.98695, 2.43401)
    (0.98905, 2.43146)
    (0.99035, 2.25433)
    (0.99145, 2.25033)
    (0.99265, 2.20755)
    (0.9934, 2.12449)
    (0.9941, 2.0745)
    (0.9945, 2.04474)
    (0.9949, 1.97179)
    (0.99515, 1.92578)
    (0.9954, 1.86836)
    (0.9954, 1.78251)
};
\end{axis}
\end{tikzpicture}\hspace{2mm}
}
\subfloat[OpenAI-1536]{\vspace{-2mm}
\begin{tikzpicture}[scale=1]
\begin{axis}[
    height=\columnwidth/2.70,
width=\columnwidth/1.90,
xlabel=recall@20,
ylabel=Qpsx100,
label style={font=\scriptsize},
tick label style={font=\scriptsize},
ymajorgrids=true,
xmajorgrids=true,
grid style=dashed,
]
\addplot[line width=0.15mm,color=amaranth,mark=o,mark size=0.5mm]
plot coordinates {
    ( 0.863, 8.281 )
    ( 0.908, 7.151 )
    ( 0.93, 6.28 )
    ( 0.945, 5.642 )
    ( 0.953, 5.056 )
    ( 0.961, 4.585 )
    ( 0.966, 3.214 )
    ( 0.97, 3.06 )
    ( 0.973, 3.144 )
    ( 0.975, 2.949 )
    ( 0.977, 2.826 )
    ( 0.979, 2.673 )
    ( 0.981, 2.53 )
    ( 0.983, 2.407 )
    ( 0.984, 2.293 )
    ( 0.986, 2.19 )
    ( 0.987, 2.098 )
    ( 0.988, 2.013 )
    ( 0.989, 1.862 )
};
\addplot[line width=0.15mm,color=amber,mark=triangle,mark size=0.5mm]
plot coordinates {
    ( 0.864, 14.138 )
    ( 0.911, 12.577 )
    ( 0.933, 11.395 )
    ( 0.947, 10.502 )
    ( 0.956, 9.79 )
    ( 0.963, 9.107 )
    ( 0.967, 8.547 )
    ( 0.971, 8.067 )
    ( 0.974, 7.628 )
    ( 0.977, 7.291 )
    ( 0.978, 6.915 )
    ( 0.981, 6.612 )
    ( 0.982, 6.31 )
    ( 0.984, 6.042 )
    ( 0.985, 5.822 )
    ( 0.987, 5.59 )
    ( 0.988, 5.383 )
    ( 0.989, 5.015 )
};
\addplot[line width=0.15mm,color=navy,mark=oplus,mark size=0.5mm]
plot coordinates {
    ( 0.74865, 9.864 )
    ( 0.81145, 8.753 )
    ( 0.86425, 6.765 )
    ( 0.90665, 5.147 )
    ( 0.9419, 3.779 )
    ( 0.9638, 2.352 )
    ( 0.9769, 1.256 )
};
\addplot[line width=0.15mm,color=black,mark=square,mark size=0.5mm]
plot coordinates {
    ( 0.7292, 18.921440824788688 )
    ( 0.7334, 10.591119729511593 )
    ( 0.7355, 5.82327195925734 )
};
\addplot[line width=0.15mm,color=black,mark=x,mark size=0.5mm]
plot coordinates {
(0.8521, 1.95899)
(0.89705, 1.41472)
(0.9203, 0.989135)
(0.9349, 0.770396)
(0.9456, 0.628691)
(0.95285, 0.528504)
(0.95815, 0.461514)
(0.9626, 0.40577)
(0.9666, 0.362753)
(0.97005, 0.327993)
(0.9731, 0.299387)
(0.9759, 0.275204)
(0.97745, 0.254601)
(0.97955, 0.236552)
(0.9806, 0.221066)
(0.9822, 0.207801)
(0.9834, 0.195677)
(0.9844, 0.185122)
(0.9853, 0.175716)
(0.98625, 0.166901)
};
\addplot[line width=0.15mm,color=violate,mark=diamond,mark size=0.5mm]
plot coordinates {
    (0.90405, 6.49527)
    (0.92645, 5.65502)
    (0.94095, 5.15528)
    (0.94915, 4.6658)
    (0.95705, 4.30346)
    (0.962, 3.97174)
    (0.96605, 3.65288)
    (0.96965, 3.42822)
    (0.9719, 3.18758)
    (0.9735, 3.00507)
    (0.97545, 2.83902)
    (0.9774, 2.68656)
    (0.97925, 2.56819)
    (0.98055, 2.4395)
    (0.98185, 2.31591)
    (0.9829, 2.20876)
    (0.984, 2.12581)
    (0.98455, 2.04137)
    (0.985, 1.95301)
};
\end{axis}
\end{tikzpicture}\hspace{2mm}
}
\subfloat[OpenAI-3072]{\vspace{-2mm}
\begin{tikzpicture}[scale=1]
\begin{axis}[
    height=\columnwidth/2.70,
width=\columnwidth/1.90,
xlabel=recall@20,
ylabel=Qpsx100,
label style={font=\scriptsize},
tick label style={font=\scriptsize},
ymajorgrids=true,
xmajorgrids=true,
grid style=dashed,
]
\addplot[line width=0.15mm,color=amaranth,mark=o,mark size=0.5mm]
plot coordinates {
    ( 0.85, 2.894 )
    ( 0.893, 2.568 )
    ( 0.919, 2.314 )
    ( 0.935, 2.112 )
    ( 0.944, 1.95 )
    ( 0.952, 1.808 )
    ( 0.957, 1.69 )
    ( 0.963, 1.588 )
    ( 0.966, 1.497 )
    ( 0.969, 1.417 )
    ( 0.971, 1.345 )
    ( 0.973, 1.28 )
    ( 0.976, 1.219 )
    ( 0.977, 1.166 )
    ( 0.979, 1.119 )
    ( 0.98, 1.077 )
    ( 0.982, 1.035 )
    ( 0.983, 0.997 )
    ( 0.985, 0.929 )
};
\addplot[line width=0.15mm,color=amber,mark=triangle,mark size=0.5mm]
plot coordinates {
    ( 0.851, 6.905 )
    ( 0.895, 6.444 )
    ( 0.919, 6.12 )
    ( 0.935, 5.818 )
    ( 0.945, 5.555 )
    ( 0.952, 5.356 )
    ( 0.957, 5.149 )
    ( 0.961, 4.93 )
    ( 0.964, 4.787 )
    ( 0.967, 4.639 )
    ( 0.97, 4.489 )
    ( 0.972, 4.333 )
    ( 0.973, 4.218 )
    ( 0.975, 4.101 )
    ( 0.977, 3.974 )
    ( 0.978, 3.867 )
    ( 0.979, 3.768 )
    ( 0.98, 3.667 )
    ( 0.982, 2.985 )
};
\addplot[line width=0.15mm,color=navy,mark=oplus,mark size=0.5mm]
plot coordinates {
    ( 0.73715, 7.048 )
    ( 0.8009, 6.560 )
    ( 0.8503, 5.803 )
    ( 0.8917, 4.680 )
    ( 0.91915, 3.451 )
    ( 0.9351, 2.274 )
    ( 0.9426, 1.342 )
};
\addplot[line width=0.15mm,color=black,mark=square,mark size=0.5mm]
plot coordinates {
    ( 0.7248, 15.229853479593994 )
    ( 0.7358, 8.306562514487463 )
    ( 0.74165, 4.494948474869056 )
    ( 0.7443, 2.409659918319539 )
};
\addplot[line width=0.15mm,color=black,mark=x,mark size=0.5mm]
plot coordinates {
(0.8374, 0.918597)
(0.88365, 0.624975)
(0.9066, 0.438494)
(0.92065, 0.342155)
(0.9318, 0.277012)
(0.9408, 0.233876)
(0.94775, 0.201733)
(0.95275, 0.177513)
(0.9566, 0.158054)
(0.9603, 0.142929)
(0.96265, 0.129939)
(0.9651, 0.119502)
(0.96765, 0.110464)
(0.97, 0.102819)
(0.97215, 0.0959188)
(0.97345, 0.0900797)
(0.9754, 0.0848493)
(0.9773, 0.0802391)
(0.97855, 0.076064)
(0.9795, 0.0723648)
};
\addplot[line width=0.15mm,color=violate,mark=diamond,mark size=0.5mm]
plot coordinates {
(0.88385, 2.48025)
(0.9093, 1.76464)
(0.9243, 1.70075)
(0.93345, 1.97375)
(0.94025, 1.83487)
(0.94565, 1.71249)
(0.9504, 1.60783)
(0.9535, 1.50734)
(0.9564, 1.43289)
(0.9587, 1.34991)
(0.96045, 1.28018)
(0.9623, 1.21308)
(0.96405, 1.16183)
(0.96545, 1.11681)
(0.96655, 1.07143)
(0.968, 1.02522)
(0.969, 0.988756)
(0.96995, 0.953474)
(0.97075, 0.9201)
};
\end{axis}
\end{tikzpicture}\hspace{2mm}
}
\\
\subfloat[DEEP]{\vspace{-2mm}
\begin{tikzpicture}[scale=1]
\begin{axis}[
    height=\columnwidth/2.70,
width=\columnwidth/1.90,
xlabel=recall@20,
ylabel=Qpsx100,
label style={font=\scriptsize},
tick label style={font=\scriptsize},
ymajorgrids=true,
xmajorgrids=true,
grid style=dashed,
]
\addplot[line width=0.15mm,color=amaranth,mark=o,mark size=0.5mm]
plot coordinates {
(0.7881, 47.3608)
(0.8797, 42.1622)
(0.91965, 36.4016)
(0.9421, 34.1142)
(0.9543, 32.3877)
(0.9628, 29.7929)
(0.97065, 28.6572)
(0.9761, 25.8645)
(0.9803, 24.7353)
(0.98295, 23.128)
(0.98515, 22.1476)
(0.98725, 20.9815)
(0.98865, 19.7699)
(0.98985, 19.4647)
(0.991, 18.703)
(0.9919, 18.2366)
(0.99275, 17.8507)
(0.9934, 17.2581)
(0.99355, 13.515)
(0.99385, 10.5818)
};
\addplot[line width=0.15mm,color=amber,mark=triangle,mark size=0.5mm]
plot coordinates {
    ( 0.788, 68.22 )
    ( 0.88, 54.552 )
    ( 0.92, 47.58 )
    ( 0.943, 42.599 )
    ( 0.955, 38.867 )
    ( 0.964, 36.305 )
    ( 0.97, 33.619 )
    ( 0.976, 31.535 )
    ( 0.979, 29.867 )
    ( 0.982, 28.383 )
    ( 0.984, 27.152 )
    ( 0.985, 25.821 )
    ( 0.987, 24.877 )
    ( 0.988, 23.819 )
    ( 0.99, 23.023 )
    ( 0.99, 22.304 )
    ( 0.992, 20.934 )
};
\addplot[line width=0.15mm,color=navy,mark=oplus,mark size=0.5mm]
plot coordinates {
    ( 0.78625, 34.155 )
    ( 0.8797, 26.317 )
    ( 0.9411, 18.792 )
    ( 0.9766, 12.230 )
    ( 0.9916, 7.290 )
    ( 0.9982, 4.067 )
};
\addplot[line width=0.15mm,color=black,mark=square,mark size=0.5mm]
plot coordinates {
    ( 0.5329, 28.53861470923145 )
    ( 0.5434, 15.868462682142163 )
    ( 0.5478, 7.575590069121401 )
    ( 0.549, 4.458573970520307 )
    ( 0.54905, 2.3176880983536037 )
};
\addplot[line width=0.15mm,color=black,mark=x,mark size=0.5mm]
plot coordinates {
(0.84545, 13.8642)
(0.88985, 9.7575)
(0.91825, 7.0884)
(0.935, 6.05219)
(0.9478, 5.22246)
(0.957, 4.53051)
(0.96285, 4.07735)
(0.9679, 3.64463)
(0.97115, 3.21336)
(0.97525, 3.06001)
(0.9779, 2.79994)
(0.98015, 2.66777)
(0.98275, 2.48735)
(0.98455, 2.34504)
(0.986, 2.21118)
(0.98755, 2.03856)
(0.98855, 1.9624)
(0.98955, 1.87689)
(0.99035, 1.82979)
};
\addplot[line width=0.15mm,color=violate,mark=diamond,mark size=0.5mm]
plot coordinates {
(0.8754, 41.3104)
(0.9143, 36.9414)
(0.93655, 34.1818)
(0.94905, 32.0298)
(0.95695, 28.5727)
(0.9647, 26.1151)
(0.9698, 24.869)
(0.9742, 21.8909)
(0.97635, 21.99)
(0.9785, 21.1881)
(0.98035, 19.8299)
(0.9818, 18.934)
(0.9832, 18.4331)
(0.98445, 17.8095)
(0.9852, 16.5409)
(0.986, 16.5115)
(0.98645, 15.835)
(0.9869, 14.8325)
(0.98715, 14.2403)
};
\end{axis}
\end{tikzpicture}\hspace{2mm}
}
\subfloat[WORD2VEC]{\vspace{-2mm}
\begin{tikzpicture}[scale=1]
\begin{axis}[
    height=\columnwidth/2.70,
width=\columnwidth/1.90,
xlabel=recall@20,
ylabel=Qpsx100,
label style={font=\scriptsize},
tick label style={font=\scriptsize},
ymajorgrids=true,
xmajorgrids=true,
grid style=dashed,
]
\addplot[line width=0.15mm,color=amaranth,mark=o,mark size=0.5mm]
plot coordinates {
    ( 0.865, 28.679 )
    ( 0.945, 23.768 )
    ( 0.971, 20.541 )
    ( 0.981, 18.338 )
    ( 0.986, 15.978 )
    ( 0.99, 14.393 )
    ( 0.991, 13.721 )
    ( 0.992, 12.69 )
    ( 0.993, 11.723 )
    ( 0.994, 10.427 )
    ( 0.995, 8.437 )
};
\addplot[line width=0.15mm,color=amber,mark=triangle,mark size=0.5mm]
plot coordinates {
    ( 0.858, 30.879 )
    ( 0.945, 24.133 )
    ( 0.97, 20.56 )
    ( 0.982, 18.236 )
    ( 0.987, 16.539 )
    ( 0.991, 15.323 )
    ( 0.993, 14.086 )
    ( 0.994, 13.026 )
    ( 0.995, 12.178 )
    ( 0.997, 10.65 )
    ( 0.998, 8.948 )
};
\addplot[line width=0.15mm,color=navy,mark=oplus,mark size=0.5mm]
plot coordinates {
    ( 0.7652, 35.183 )
    ( 0.8753, 29.379 )
    ( 0.9291, 20.716 )
    ( 0.9519, 14.239 )
    ( 0.95895, 8.540 )
    ( 0.9615, 4.984 )
};
\addplot[line width=0.15mm,color=black,mark=square,mark size=0.5mm]
plot coordinates {
    ( 0.39515, 13.068673544104267 )
    ( 0.4026, 6.486415199032966 )
    ( 0.40485, 3.1889706699212064 )
    ( 0.4059, 1.5994824888481565 )
    ( 0.40615, 0.8148235602194852 )
    ( 0.4062, 0.4561623495705986 )
};
\addplot[line width=0.15mm,color=black,mark=x,mark size=0.5mm]
plot coordinates {
(0.8576, 3.42164)
(0.94175, 4.03645)
(0.96785, 2.67131)
(0.97985, 2.04611)
(0.98555, 1.65488)
(0.98975, 1.3923)
(0.99235, 1.20094)
(0.9938, 1.06832)
(0.99535, 0.95318)
(0.99635, 0.869266)
(0.9967, 0.79894)
(0.99715, 0.742706)
(0.99735, 0.690741)
(0.99775, 0.649583)
(0.99795, 0.614293)
(0.99805, 0.582423)
(0.9983, 0.554952)
(0.9985, 0.530866)
(0.9986, 0.510026)
(0.9988, 0.492162)
};
\addplot[line width=0.15mm,color=violate,mark=diamond,mark size=0.5mm]
plot coordinates {
    (0.9443, 19.2348)
    (0.96935, 16.1854)
    (0.9813, 13.5796)
    (0.98685, 11.6955)
    (0.9904, 10.595)
    (0.9925, 9.6672)
    (0.9936, 9.08364)
    (0.99485, 8.46861)
    (0.99565, 7.83637)
    (0.9961, 7.47134)
    (0.99625, 7.12595)
    (0.99645, 6.67925)
    (0.99675, 6.42542)
    (0.99695, 6.21296)
    (0.99715, 5.8821)
    (0.9973, 5.68706)
    (0.99745, 5.42351)
    (0.9976, 5.27949)
    (0.9977, 5.08007)
};
\end{axis}
\end{tikzpicture}\hspace{2mm}

}
\subfloat[MSMARCO10M]{\vspace{-2mm}
\begin{tikzpicture}[scale=1]
\begin{axis}[
    height=\columnwidth/2.70,
width=\columnwidth/1.90,
xlabel=recall@20,
ylabel=Qpsx100,
label style={font=\scriptsize},
tick label style={font=\scriptsize},
ymajorgrids=true,
xmajorgrids=true,
grid style=dashed,
]
\addplot[line width=0.15mm,color=amaranth,mark=o,mark size=0.5mm]
plot coordinates {
    ( 0.921, 6.375 )
    ( 0.956, 3.867 )
    ( 0.969, 2.854 )
    ( 0.976, 2.287 )
    ( 0.981, 1.864 )
    ( 0.985, 1.595 )
    ( 0.988, 1.398 )
    ( 0.989, 1.235 )
    ( 0.991, 1.119 )
    ( 0.992, 1.019 )
    ( 0.994, 0.863 )
    ( 0.995, 0.746 )
    ( 0.996, 0.658 )
    ( 0.997, 0.534 )
};
\addplot[line width=0.15mm,color=amber,mark=triangle,mark size=0.5mm]
plot coordinates {
    ( 0.948, 3.914 )
    ( 0.973, 2.741 )
    ( 0.983, 2.314 )
    ( 0.987, 1.987 )
    ( 0.99, 1.718 )
    ( 0.993, 1.543 )
    ( 0.994, 1.41 )
    ( 0.997, 1.105 )
    ( 0.998, 0.968 )
};
\addplot[line width=0.15mm,color=navy,mark=oplus,mark size=0.5mm]
plot coordinates {
    ( 0.94255, 7.713 )
    ( 0.96695, 4.802 )
    ( 0.97995, 2.777 )
    ( 0.98855, 1.518 )
    ( 0.9934, 0.799 )
};
\addplot[line width=0.15mm,color=black,mark=square,mark size=0.5mm]
plot coordinates {
    ( 0.74905, 1.6171910490231087 )
    ( 0.76805, 0.8179127918826627 )
    ( 0.7792, 0.42581448690918215 )
    ( 0.7856, 0.21461068475635116 )
    ( 0.78845, 0.10864601474557935 )
    ( 0.79095, 0.05512706388612434 )
};
\addplot[line width=0.15mm,color=black,mark=x,mark size=0.5mm]
plot coordinates {
(0.9237, 0.629621)
(0.95655, 0.376468)
(0.9684, 0.267861)
(0.9762, 0.213027)
(0.9799, 0.176275)
(0.98305, 0.151088)
(0.98535, 0.132614)
(0.98685, 0.118718)
(0.9879, 0.10746)
(0.98855, 0.098329)
};
\addplot[line width=0.15mm,color=violate,mark=diamond,mark size=0.5mm]
plot coordinates {
    (0.93175, 2.96667)
    (0.96065, 3.0461)
    (0.9718, 2.13863)
    (0.9778, 2.11502)
    (0.98065, 1.75713)
    (0.98345, 1.48583)
    (0.98495, 1.2937)
    (0.98625, 1.12832)
    (0.9874, 1.03176)
    (0.98885, 0.937897)
    (0.98945, 0.850725)
    (0.9901, 0.795442)
    (0.99035, 0.738426)
    (0.991, 0.68375)
    (0.99125, 0.647915)
    (0.9917, 0.607636)
    (0.99215, 0.572496)
    (0.9925, 0.545818)
    (0.9929, 0.517942)
    (0.99325, 0.491956)
};
\end{axis}
\end{tikzpicture}\hspace{2mm}
}
\subfloat[TINY]{\vspace{-2mm}
\begin{tikzpicture}[scale=1]
\begin{axis}[
    height=\columnwidth/2.70,
width=\columnwidth/1.90,
xlabel=recall@20,
ylabel=Qpsx100,
label style={font=\scriptsize},
tick label style={font=\scriptsize},
ymajorgrids=true,
xmajorgrids=true,
grid style=dashed,
]
\addplot[line width=0.15mm,color=amaranth,mark=o,mark size=0.5mm]
plot coordinates {
    ( 0.773, 10.411 )
    ( 0.871, 9.966 )
    ( 0.913, 8.395 )
    ( 0.936, 7.079 )
    ( 0.951, 5.921 )
    ( 0.961, 5.202 )
    ( 0.97, 4.716 )
    ( 0.976, 4.355 )
    ( 0.98, 3.955 )
    ( 0.983, 3.629 )
    ( 0.986, 3.318 )
    ( 0.988, 3.077 )
    ( 0.99, 2.876 )
    ( 0.992, 2.695 )
    ( 0.993, 2.546 )
    ( 0.994, 2.287 )
    ( 0.995, 2.072 )
};
\addplot[line width=0.15mm,color=amber,mark=triangle,mark size=0.5mm]
plot coordinates {
    ( 0.775, 19.252 )
    ( 0.868, 13.605 )
    ( 0.911, 10.939 )
    ( 0.936, 9.375 )
    ( 0.951, 8.206 )
    ( 0.962, 6.077 )
    ( 0.97, 6.634 )
    ( 0.976, 6.112 )
    ( 0.98, 5.655 )
    ( 0.984, 5.282 )
    ( 0.987, 4.99 )
    ( 0.989, 4.715 )
    ( 0.991, 4.471 )
    ( 0.992, 4.258 )
    ( 0.993, 4.054 )
    ( 0.994, 3.888 )
    ( 0.995, 3.588 )
};
\addplot[line width=0.15mm,color=navy,mark=oplus,mark size=0.5mm]
plot coordinates {
    ( 0.70345, 19.613 )
    ( 0.82115, 14.913 )
    ( 0.9088, 10.232 )
    ( 0.95975, 6.384 )
    ( 0.98305, 3.714 )
    ( 0.99005, 2.063 )
};
\addplot[line width=0.15mm,color=black,mark=square,mark size=0.5mm]
plot coordinates {
    ( 0.42985, 6.373150343838645 )
    ( 0.457, 3.251456635014677 )
    ( 0.4737, 1.6572340500733543 )
    ( 0.48085, 0.7328861149639863 )
    ( 0.48365, 0.4191070014932105 )
    ( 0.48375, 0.2214967374246345 )
};
\addplot[line width=0.15mm,color=black,mark=x,mark size=0.5mm]
plot coordinates {
(0.6935, 1.72037)
(0.79985, 1.1025)
(0.8601, 0.758559)
(0.8955, 0.580031)
(0.9165, 0.469831)
(0.93365, 0.395489)
(0.94515, 0.341836)
(0.9547, 0.302487)
(0.96145, 0.269494)
(0.96815, 0.244272)
(0.97225, 0.222127)
(0.9763, 0.204572)
(0.97985, 0.189626)
(0.98255, 0.176256)
(0.98485, 0.165312)
(0.9868, 0.155173)
(0.98825, 0.146563)
(0.9893, 0.138324)
(0.9908, 0.131433)
(0.99195, 0.125315)
};
\addplot[line width=0.15mm,color=violate,mark=diamond,mark size=0.5mm]
plot coordinates {
(0.76645, 4.98094)
(0.8634, 7.665)
(0.90515, 5.17134)
(0.9277, 6.192)
(0.9415, 5.42559)
(0.9514, 3.85602)
(0.9596, 4.22595)
(0.9653, 4.06331)
(0.9691, 3.66579)
(0.97225, 3.33089)
(0.9752, 3.06613)
(0.97715, 2.91836)
(0.97905, 2.71688)
(0.9802, 2.53089)
(0.981, 2.37384)
(0.98145, 2.28123)
(0.98255, 2.15778)
(0.983, 2.04867)
(0.9835, 1.9653)
(0.984, 1.88247)
};
\end{axis}
\end{tikzpicture}\hspace{2mm}
}

\vgap\caption{The Test of Additional Quantization Methods}\vgap\label{fig:extra-quantiztaion}
\end{small}
\end{figure*}